\documentclass[review,fleqn,sort&compress]{elsarticle}
\usepackage{amssymb,amsbsy,amsthm,amsmath,amsfonts,amssymb,amscd}
\usepackage{geometry}
\usepackage{babel,blindtext}
\usepackage{bm} 

\usepackage{graphicx}
\usepackage{float}
\usepackage{multirow}
\usepackage{subfigure}
\usepackage{lineno}
\usepackage{hyperref}
\hypersetup{colorlinks, citecolor=black, filecolor=black, linkcolor=black, urlcolor=black}
\usepackage{booktabs}
\usepackage[ruled,linesnumbered]{algorithm2e}
\usepackage{appendix}
\usepackage{algpseudocode}
\usepackage{mathtools,nccmath}
\usepackage{adjustbox}
\usepackage{cleveref}
\usepackage{dutchcal}
\usepackage[utf8]{inputenc}
\usepackage[scr=rsfso,cal=euler,frak=euler,bb=ams]{mathalfa}
\usepackage{longtable}
\usepackage{comment}
\usepackage{color}
\usepackage[dvipsnames]{xcolor}
\usepackage{cancel}
\usepackage{soul}
\colorlet{soulred}{red!40}
\renewcommand{\arraystretch}{1.4}
\usepackage{xparse}
\usepackage{mathrsfs}
\usepackage{upgreek}
\usepackage{colortbl}
\usepackage{hhline}
\newcolumntype{C}[1]{>{\centering\let\newline\\\arraybackslash\hspace{0pt}}m{#1}}

\begin{document}

\begin{frontmatter}
\title{I-FENN with Temporal Convolutional Networks: expediting the load-history analysis of non-local gradient damage propagation}

\author{Panos Pantidis\corref{cor1}}
\author{Habiba Eldababy}
\author{Diab Abueidda}
\author{Mostafa E. Mobasher\corref{cor2}}
\cortext[cor1]{Corresponding author. \emph{E-mail address:} \texttt{pp2624@nyu.edu} (Panos Pantidis)}
\cortext[cor2]{Corresponding author. \emph{E-mail address:} \texttt{mostafa.mobasher@nyu.edu} (Mostafa Mobasher)}
\address{Civil and Urban Engineering Department, New York University Abu Dhabi, Abu Dhabi, P.O. Box 129188, UAE}

\begin{highlights}

\item A TCN-based formulation of I-FENN is developed and applied for the first time across a load-history analysis with non-local gradient damage 

\item I-FENN computational savings scale with model size, and we show a significant speedup compared to classical FEM monolithic and staggered schemes.

\item We showcase the robustness of I-FENN by satisfying very strict convergence criteria across the entire simulation 

\item We demonstrate the appropriate use of input normalization and output un-normalization when training a physics-based network with different length scales

\item I-FENN is always applied in different mesh idealizations from the one used for the TCN training  

\end{highlights}

\begin{abstract}

In this paper, we demonstrate for the first time how the Integrated Finite Element Neural Network (I-FENN) framework, previously proposed by the authors \cite{pantidis2023integrated,pantidis2023116160}, can efficiently simulate the entire loading history of non-local gradient damage propagation. To achieve this goal, we first adopt a Temporal Convolutional Network (TCN) as the neural network of choice to capture the history-dependent evolution of the non-local strain in a coarsely meshed domain. The quality of the network predictions governs the computational performance of I-FENN, and therefore we perform an extended investigation aimed at enhancing them. We explore a data-driven vs. physics-informed TCN setup to arrive at an optimum network training, evaluating the network based on a coherent set of relevant performance metrics. We address the crucial issue of training a physics-informed network with input data that span vastly different length scales by proposing a systematic way of input normalization and output un-normalization. We then integrate the trained TCN within the nonlinear iterative FEM solver and apply I-FENN to simulate the damage propagation analysis. I-FENN is always applied in mesh idealizations different from the one used for the TCN training, showcasing the framework's ability to be used at progressively refined mesh resolutions. We illustrate several cases that I-FENN completes the simulation using either a modified or a full Newton-Raphson scheme, and we showcase its computational savings compared to both the classical monolithic and staggered FEM solvers. We underline that we satisfy very strict convergence criteria for every increment across the entire simulation, providing clear evidence of the robustness and accuracy of I-FENN. All the code and data used in this work will be made publicly available upon publication of the article.
\end{abstract}

\begin{keyword}
\texttt IFENN \sep non-local damage \sep neural networks \sep finite element analysis \sep multi-physics \sep scientific computing 
\end{keyword}

\end{frontmatter}


\newpage
\section{Introduction}
\label{Section:Introduction}

\subsection{Literature review}

The vast majority of real-world problems in solid mechanics often possess a highly non-linear, history-dependent, and coupled-field nature. Material damage is one example of such a complex physical process, during which micro-cracks and micro-voids nucleate and grow to form macro-scale damage zones representing fracture inside the domain \cite{kachanovbook, lemaitrebook}. This process is phenomenologically represented by the gradual degradation of the material mechanical properties and may have catastrophic consequences on the mechanical integrity. Hence, it is a prerequisite for the development of reliable predictive tools. At the same time, it also presents formidable challenges since it requires the solution of partial differential equations (PDEs) in large and often irregular domains that exhibit strain-softening over time.  

The challenges associated with modeling fracture and damage have geared the development of a plethora of numerical approaches over the years, which have been widely adopted by both academic researchers and industrial practitioners. Indicatively, we mention the finite element method (FEM) \cite{hughes2012finite, kattan2012damage, mobasher2021dual}, extended finite element method (XFEM) \cite{moes2002extended, waisman2010detection}, boundary element method (BEM) \cite{sfantos2007multi, zhang2019nonlocal}, virtual element method (VEM) \cite{de2018virtual, liu2023virtual}, meshless approaches \cite{liu1999multiple, farahani2016extending}, and more. FEM in particular is considered one of the most robust approaches for solving damage mechanics problems across a wide range of fields, such as structural engineering \cite{crouch2013experimental}, poro-mechanics and hydraulic fracturing \cite{mobasher2021non, shauer2019improved} and bio-engineering \cite{hamed2013multiscale}. Matured over the span of several decades, FEM owes its reliability to the diligent exploration of its fundamental principles: convergence performance \cite{binev2004adaptive, ciarlet1973maximum}, integration laws \cite{malkus1978mixed}, stability conditions \cite{brezzi1990discourse} and other properties \cite{nguyen2008smooth}. Indisputably, the major bottleneck of the finite element method is the notorious computational cost, which scales with the size of the investigated problem and may pose a limiting factor for large-scale problems.    

More recently, the exponential growth in the field of machine learning (ML) has attracted the attention of the engineering sector, and ML-based methods have been investigated as alternatives to more established numerical or analytical methods in computational mechanics \cite{saha2021hierarchical, liu2021knowledge, samaniego2020energy, oishi2017computational, masi2021thermodynamics, mozaffar2019deep}. Broadly, there are two distinctive ways in which ML-based algorithms can be used for such problems: the data-driven \cite{kirchdoerfer2016data} and the physics-informed approach \cite{karniadakis2021physics}. Data-driven models use labeled data pairs to interpret the relationship between input and output quantities, relying entirely on existing datasets. A classic example of this approach is the work by Ortiz and collaborators \cite{kirchdoerfer2016data, karapiperis2021data, carrara2020data}, in which case material-point computations are carried out directly from experimental data and bypass the empirical material modeling. For physics-based models, prior physical knowledge is infused into the network objective function such as the input variables and the resulting predictions satisfy a known physical expression \cite{cuomo2022scientific}. These models are often used as surrogate approximates of the solution of PDEs in several mechanics problems, such as heat transfer \cite{cai2021physics}, Navier-Stokes equations \cite{jin2021nsfnets}, stress analysis \cite{thakur2022viscoelasticnet}, and coupled-field problems \cite{haghighat2022physics}. Overall, the intersection between the computational mechanics and machine learning fields is constantly growing, and the interested reader is referred to \cite{herrmann2024} for a more detailed survey of the existing literature.  


While ML models have generally shown some tangible and promising results in the field of computational mechanics, they suffer from fundamental shortcomings that currently prevent them from becoming rigorous and standalone alternative solvers. For example, a major drawback of data-driven models is their limited interpretability, which is a subject of intense ongoing research \cite{sun2022data, manfren2022data}. Another well-known deficiency of these models is their poor generalization capability across datasets which are different from the training one \cite{fuhg2022physics}. On the other hand, physics-based models may often fail to learn more complex tasks, consequently yielding inaccurate predictions \cite{pantidis2023116160, daw2022rethinking}. In both cases, there are still several major open questions regarding the network convergence performance \cite{shin2020convergence, mishra2022estimates}, training stability \cite{krishnapriyan2021characterizing, wang2021understanding} and sensitivity to hyperparameters \cite{pantidis2023116160, markidis2021old}. The aforementioned issues pose major challenges and hinder the direct adoption of these models in real-world engineering applications, where accuracy and reliability are of utmost importance.

In view of the above, there is a true need to develop frameworks of a hybrid nature, where the exponentially growing expressivity of ML-based algorithms could be combined with the long-standing robustness of FEM. This research gap is still far from being considered complete, but it has already sparked a few studies in this direction. Mitusch et al. \cite{mitusch2021hybridFEMNN} presented such a hybrid framework for inverse problems, where PDEs were first augmented with neural networks to represent unknown terms and subsequently discretized in space with FEM. The FENA framework is another approach \cite{jokar2022two}, where pre-trained neural networks are utilized as surrogate building blocks of the physical system. Another hybrid methodology was developed in \cite{garcia2020machine} for the multi-scale stress computation in geological sub-surfaces, where FEM and NNs are applied at different mesh levels to obtain different parts of the final solution. Overall, however, we underline that these studies are still sparse in the literature, and even more importantly, these frameworks were not examined against non-linear and path-dependent problems. Therefore, their applicability to these substantially more challenging problems has yet to be demonstrated.

\subsection{Scope and Outline}
\label{Scope_and_Outline}

Motivated by these gaps, in \cite{pantidis2023integrated}, we proposed a framework that couples FEM with NNs in a conceptually different way than all the aforementioned hybrid schemes. This new framework is termed Integrated Finite Element Neural Network (I-FENN) and aims to accelerate the solution of non-linear multi-physics problems in solid mechanics. The core idea of I-FENN is to split the governing PDEs into two categories: a) the mechanical equilibrium condition is satisfied through FEM numerical discretization, while b) a pre-trained network is used to solve a physics-related PDE that is coupled with the equilibrium condition. The key to coupling the two numerical methods is to invoke the pre-trained network inside the element stiffness function in a fashion that resembles any user-defined material-level functionality. This approach allows the displacements to be treated only as nodal unknowns. At the same time, the other state variable is computed at the integration points based on the pre-trained network and the displacement field. Therefore, the computational gains stem from solving a smaller system of equations while preserving the knowledge of the physics-related state variable at the integration points. We also note that we maintain the incremental and iterative nature of the FEM solver, which ensures both the convergence of the solution as well as the continuous update of the physics-related state variable.  

I-FENN has so far applied in the cases of non-local gradient damage \cite{pantidis2023integrated, pantidis2023116160} and fully coupled transient thermo-elasticity \cite{abueidda2023fenn}. In the case of gradient damage, our previous studies focused on developing the fundamental principles of the framework \cite{pantidis2023integrated} and examining the network convergence and hyperparameter impact \cite{pantidis2023116160}. However, the feasibility of this concept has so far been shown only at individual load increments. It is yet to be demonstrated if and how I-FENN can successfully utilize the outputs of one increment as input for the next one, simulating the history of damage growth. Also, there remains an open question of whether computational savings are attainable. These are the fundamental gaps that we address in this work.  

The need for modeling the load history of damage propagation points to the adoption of sequence-to-sequence models. These networks treat a $sequence$ of data as input and output quantities. Several candidate models exist, such as Recurrent Neural Networks (RNNs) and its extensions GRUs and LSTMs \cite{mozaffar2019deep}, Transformer-based models \cite{vaswani2017attention} and Temporal Convolutional Networks (TCNs) \cite{bai2018empirical}. At varying levels of computational efficiency and accuracy, all the above have shown successful signs in performing sequential tasks in engineering mechanics \cite{cuomo2022scientific, wu2022recurrent, abueidda2023fenn}. TCNs have shown to be more advantageous than RNNs when handling long-term dependencies \cite{aksan2019stcn}, because their non-recurrent structure allows for consistent gradients propagation across time steps and reduces the risk of exploding or vanishing gradients. Also, a peculiarity of the damage propagation problem of interest is that each time step has to be informed only from the past and current time steps. Transformers require an additional step of proper identification and labeling of the time steps in order to achieve this \cite{zhongbo2024pre}, whereas TCNs inherently alleviate this requirement. Therefore, here we select the TCN architecture and a more detailed justification of their relevance and suitability is provided in Section \ref{Section:TCN_relevance_architecture}.

Departing from our previous investigations, the novel contributions of this paper are summarized as follows:

\begin{itemize}

    \item We present an overarching I-FENN algorithm, along with its implementation details, with an appropriate sequence-to-sequence neural network, the Temporal Convolutional Network (TCN). 
     
    \item We demonstrate evidently and for the first time that I-FENN can successfully simulate the propagation of gradient damage by satisfying strict convergence criteria at every increment of the load history sequence.

    \item We display the computational savings that can be achieved with I-FENN and how these gains increase against progressively refined mesh resolutions.
   
    \item We examine several crucial aspects of the highly modular TCN-based I-FENN: a) full vs. modified Newton-Raphson implementation, b) data-driven vs. physics-informed training, c) different techniques for input normalization and output un-normalization, d) spatial gradient computation using shape functions.
    
\end{itemize}

The paper is structured as follows. Section \ref{Section:Problem_statement} presents a brief overview of non-local gradient damage and its numerical treatment with conventional FEM. Section \ref{Section:IFENN_framework} presents the overarching algorithm and mathematical setup of I-FENN. Section \ref{Section:TCN} discusses the fundamental details appertaining to Temporal Convolutional Networks, such as the network architecture, the data-driven vs. physics-informed setup, and the normalization \& un-normalization step of the network inputs and outputs. Section \ref{Section:IFENN_worfklow} describes the details of implementing I-FENN with TCNs for the history analysis of non-local gradient damage. Section \ref{Section:Results} presents the numerical results of this study, and we conclude with a summary of our work and outlook discussion in Section \ref{Section:Discussion_Conclusions}.  

\section{Problem statement}
\label{Section:Problem_statement}

\subsection{Non-local gradient damage}
\label{Section:Problem_statement_nonlocal_damage}

\begin{figure}
    \centering
    \includegraphics[width=0.5\textwidth]{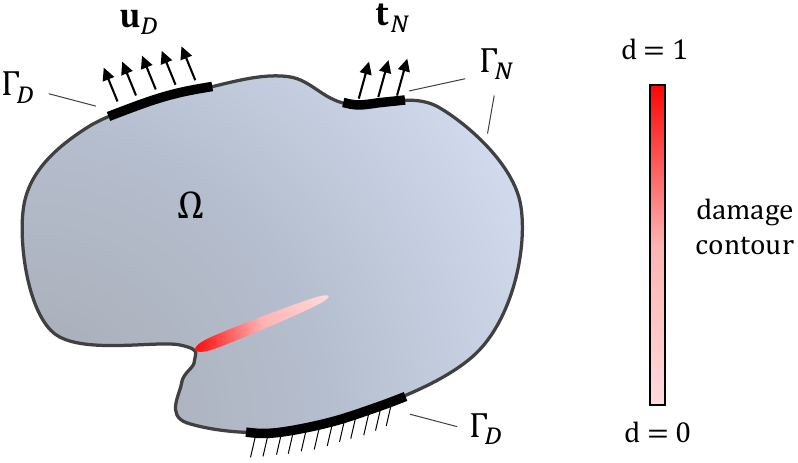}
    \caption{Schematic of an elastic domain with a sample damage contour}
    \label{Fig:Figure_schematic_domain}
\end{figure}

Let $\Omega$ be an elastic domain with a boundary $\Gamma$, as shown schematically in Figure \ref{Fig:Figure_schematic_domain}. A displacement load ${\boldsymbol{u}}_{D}$ is prescribed at the Dirichlet boundary part $\Gamma_{D}$, and traction forces ${\boldsymbol{t}}_{N}$ are applied at the Neumann boundary part $\Gamma_{N}$. Using tensor notation, let us define with $\boldsymbol{\sigma}$, $\boldsymbol{\varepsilon}$, $\boldsymbol{u}$ and $\boldsymbol{C}$ the tensors for the Cauchy stress, the strain, the displacement and the constitutive matrix respectively. Assuming small deformation theory, the governing equations for equilibrium, constitutive law, and compatibility can be expressed as:

\begin{equation}
	\boldsymbol{\nabla} \cdot \boldsymbol{\sigma} = 0 \; \; \; in \; \; \; \Omega \; ; \; \; \; \; \; \; 
    \boldsymbol{u} = {\boldsymbol{u}}_{D} \; \; \; on \; \; \; \Gamma_{D} \; ; \; \; \; \; \; \; 
    \boldsymbol{\sigma} \cdot \boldsymbol{n} = {\boldsymbol{t}}_{N} \; \; \; on \; \; \; \Gamma_{N} \;
\label{Eqn_equilibrium}
\end{equation} 

\begin{equation}
	\boldsymbol{\sigma} = \left( 1 - d \right) \boldsymbol{C} \boldsymbol{:} \boldsymbol{\varepsilon}
\label{Eqn_constitutive}
\end{equation}

\begin{equation}
	\boldsymbol{\varepsilon} = \dfrac{1}{2} \left( \boldsymbol{\nabla} \boldsymbol{u} + \left[\boldsymbol{\nabla} \boldsymbol{u} \right]^{T} \right)
\label{Eqn_compatibility}
\end{equation}

\noindent where ${\boldsymbol{\nabla}} \cdot$ is the divergence operator, ${\boldsymbol{\nabla}}$ is the gradient operator and $\boldsymbol{n}$ represents the outward unit vector of the boundary. In continuum damage mechanics theory, the smeared crack is approximated as a continuous zone with reduced stiffness. The magnitude of the stiffness reduction is dictated by a scalar variable $d$, which enters the right-hand side of the constitutive equation \ref{Eqn_constitutive} and ranges between 0 (no damage) and 1 (fully damaged state). Damage is typically computed by material-specific phenomenological laws as a function of the local equivalent strain $\varepsilon_{eq}$, which is an invariant measure of the material deformation ($d = d(\varepsilon_{eq})$). However, the local definition of damage yields well-documented spurious mesh-dependent results during the numerical solution \cite{geers1998strain}, which can be in turn alleviated if one resorts to non-local strain formulations. Among the wide range of options (see for example \cite{pijaudier1987nonlocal, moes2021lip, deborstcomparison}), here we adopt the non-local gradient-enhanced model by Peerlings et al. \cite{peerlings1996gradient} where damage is diffused over a characteristic length scale $l_{c}$. The setup of the problem can now be complemented by the following expressions:

\begin{equation}
    \bar{\varepsilon}_{eq} - g \boldsymbol{\nabla}^{2} \bar{\varepsilon}_{eq} = \varepsilon_{eq} \; \; \; in \; \; \; \Omega
\label{Eqn_NonlocalGradientPDE_1}
\end{equation}

\begin{equation}
    \boldsymbol{\nabla} \bar{\varepsilon}_{eq} \cdot \boldsymbol{n} = 0 \; \; \; on \; \; \; \Gamma
\label{Eqn_NonlocalGradientPDE_2}
\end{equation}

\noindent where Eqn. \ref{Eqn_NonlocalGradientPDE_1} describes the local-to-nonlocal strain field coupling, Eqn. \ref{Eqn_NonlocalGradientPDE_2} is the boundary condition expression, $\boldsymbol{\nabla}^{2}$ is the Laplacian operator, and $g = (l_{c})^2/2$. In this case, the damage becomes a function of the non-local equivalent strain at each material point, $d = d(\bar{\varepsilon}_{eq})$.  

\subsection{Finite element analysis}
\label{Section:Problem_statement_FEM}

For the benchmark numerical solution of this problem we adopt the Finite Element Method (FEM) \cite{hughes2012finite}. Let $\bf{N}^{u}$ and $\bf{N}^{\bar{\varepsilon}}$ be the shape function matrices of the displacements and non-local strains, and $\bf{B}^{u}$ and $\bf{B}^{\bar{\varepsilon}}$ their derivatives. Assuming a displacement-driven problem with zero traction, the discretized weak form of the displacement and strain residuals are:

\begin{equation}
    {\bf{R}}^{u} = \int_{\Omega} \left[{\bf{B}^u}\right]^{T} {\boldsymbol{\sigma}} \; d\Omega 
\label{FEM_Weak_Res_u}
\end{equation}

\begin{equation}
    {\bf{R}}^{\bar{\varepsilon}} = {\int_\Omega \left[\bf{N}^{\bar{\varepsilon}}\right]^T \boldsymbol{\bar{\varepsilon}} \; d\Omega \; + \int_\Omega \left[\bf{B}^{\bar \varepsilon}\right]^T g \nabla \boldsymbol{\bar{\varepsilon}} \; d\Omega} \; - {\int_\Omega \left[\bf{N}^{\bar{\varepsilon}}\right]^T \boldsymbol{\Tilde{\varepsilon}} \; d\Omega}
\label{FEM_Weak_Res_e}
\end{equation}

The non-linear nature of the problem dictates the minimization of the nodal residuals using an iterative solver, and here we utilize the Newton-Raphson scheme. The linearized system of equations reads as:

\begin{equation}
    \underbrace{\begin{bmatrix} {\bf{J}}^{uu} & {\bf{J}}^{u\bar{\varepsilon}} \\ {\bf{J}}^{\bar{\varepsilon}u} & {\bf{J}}^{\bar{\varepsilon}\bar{\varepsilon}}
    \end{bmatrix}}_{{\bf{J}}}
    \begin{bmatrix} {\delta \boldsymbol{u}} \\ {\delta \boldsymbol{\bar{\varepsilon}}}
    \end{bmatrix} = - 
    \begin{bmatrix} {\bf{R}}^{u} \\ {\bf{R}}^{\bar{\varepsilon}}  
    \end{bmatrix} 
    \label{FEM_SystemEqn_Full}
\end{equation}

\noindent where the entries of the Jacobian matrix $\bf{J}$ can be computed as \cite{peerlings1996gradient}:

\begin{equation}
      {\bf{J}}^{uu} = \frac{\partial {\bf{R}}^{u}}{\partial \boldsymbol{u}} = \int_{\Omega} \left[{\bf{B}^u}\right]^{T} {\boldsymbol{(1-d)}} {\boldsymbol{C}} {\bf{B}^{u}} \; d\Omega
\label{Eqn_Juu}
\end{equation}

\begin{equation}
      {\bf{J}}^{u \bar{\varepsilon}} = \frac{\partial {\bf{R}}^{u}}{\partial \boldsymbol{\bar{\varepsilon}}} = - 
      \int_{\Omega} \left[{\bf{B}^u}\right]^{T} \boldsymbol{C} \frac{\partial \boldsymbol{d}}{\partial \bar{\varepsilon}} \varepsilon {\bf{N}}^{\bar{\varepsilon}} \; d\Omega
\label{Eqn_Jue}
\end{equation}

\begin{equation}
\begin{split}
      {\bf{J}}^{\bar{\varepsilon} u} = \frac{\partial {\bf{R}}^{\bar{\varepsilon}}}{\partial \boldsymbol{u}} = - \int_{\Omega} \left[{\bf{N}^{\bar{\varepsilon}}}\right]^{T} \frac{\partial \varepsilon_{eq}}{\partial \varepsilon_{ij}} {\boldsymbol{B}^{u}} \; d\Omega
\end{split}
\label{Eqn_Jeu}
\end{equation}

\begin{equation}
\begin{split}
      {\bf{J}}^{\bar{\varepsilon} \bar{\varepsilon}} = \frac{\partial {\bf{R}}^{\bar{\varepsilon}}}{\partial \boldsymbol{\bar{\varepsilon}}} = \int_{\Omega} \left(\left[{\bf{N}^{\bar{\varepsilon}}}\right]^{T} \boldsymbol{N}^{\bar{\varepsilon}} + \left[{\bf{B}^{\bar{\varepsilon}}}\right]^{T} g {\bf{B}^{\bar{\varepsilon}}}\right) \; d\Omega
\end{split}
\label{Eqn_Jee}
\end{equation}

Each increment is considered converged at the $i^{th}$ iteration, if the relative change in the norm of the nodal degree of freedom vector between the $1^{st}$ and the $i^{th}$ iteration drops below a tolerance value $tol$: 

\begin{equation}
    r_u = \frac{\prescript{}{i}{\| \left[ \delta{\boldsymbol{u}} \; \delta \boldsymbol{\bar{\varepsilon}} \right]^{T} \|_{2}}}{\prescript{}{1}{\| \left[ \delta{\boldsymbol{u}} \; \delta \boldsymbol{\bar{\varepsilon}} \right]^{T} \|_{2}}}  < tol
\label{Eqn_convergence_criterion_FEM}
\end{equation}

\noindent where the left subscript denotes the iteration number. 

The procedure outlined above represents a monolithic numerical solution, where the incremental changes of the displacement and non-local strain degrees of freedom are computed simultaneously at each iteration. This process requires the solution of the system in Eqn. \ref{FEM_SystemEqn_Full}, which, in general, yields a fast convergence rate. In some cases, however, it can be challenging to simultaneously satisfy convergence for both displacements and non-local strains and alternatively, a staggered approach can be used. In this case, $\delta \boldsymbol{u}$ and $\delta \boldsymbol{\bar{\varepsilon}}$ are computed via an alternating minimization: the displacement and non-local strain variables are solved sequentially, with one of them being fixed while calculating the other, and vice versa. Therefore, only ${\bf{J}}^{uu}$ and ${\bf{J}}^{\varepsilon \varepsilon}$ need to be computed in this case. This method, however, requires generally more iterations than the monolithic approach and typically yields a slower convergence rate \cite{gerasimov2016line}. Since I-FENN is conceptually closer to the staggered approach, in this paper, we benchmark the I-FENN results against both the FEM-monolithic and the FEM-staggered solvers. The implementation algorithms of the two FEM solvers are provided in \ref{Appendix:FEM_Algorithms}.

\section{I-FENN framework for the load-history analysis of non-local
gradient damage}
\label{Section:IFENN_framework}

\begin{figure}[H]
    \centering
    \includegraphics[width=1\textwidth]{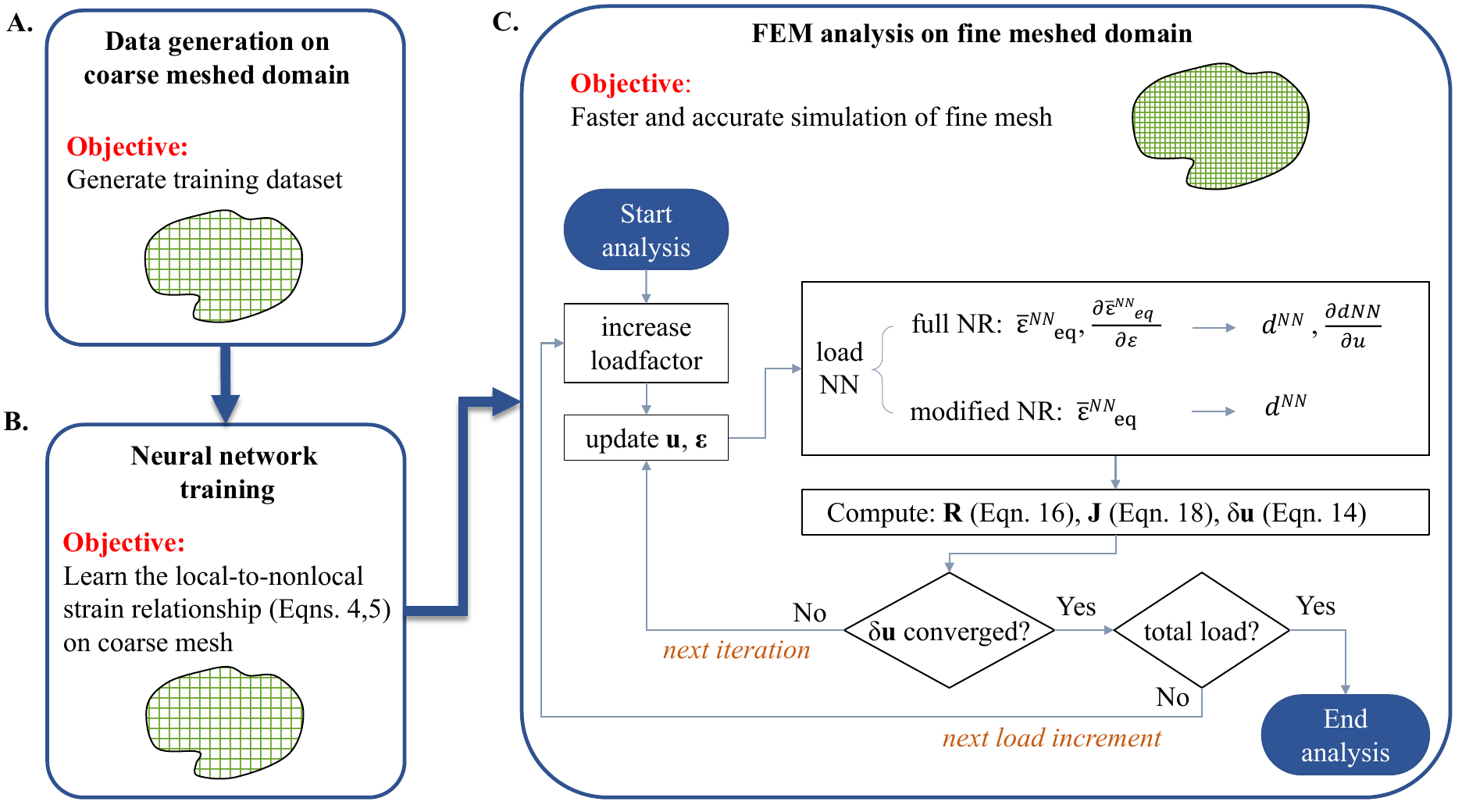}
    \caption{Schematic overview of the overarching IFENN framework.}
    \label{Fig:Figure_IFENN_Stage0}
\end{figure}

The core idea of I-FENN is that a pre-trained neural network (NN) can act as an approximator of a PDE solution, and therefore, it can be deployed in the FEM stiffness function to swiftly approximate the state variable of interest. In the case of non-local gradient damage, the queried PDE is given in Eqn. \ref{Eqn_NonlocalGradientPDE_1}. As shown in Fig. \ref{Fig:Figure_IFENN_Stage0}, the current overarching IFENN algorithm has three steps: a) an FEM analysis on a coarsely meshed model, in order to generate the network training dataset, b) the training of the neural network, which yields the local-to-non-local strain mapping function, c) an FEM analysis on a finer discretized model with the trained neural network being integrated in the FEM solver. In this section, we focus our discussion on the role of the network and on the mathematical background of I-FENN, and a detailed description of each step is provided in Section \ref{Section:IFENN_worfklow}.

The neural network targets the local-to-nonlocal strain transformation of Eqn. \ref{Eqn_NonlocalGradientPDE_1}, therefore its input variables are the coordinates $x$, $y$ and the equivalent local strain $\varepsilon_{eq}$ of the material points. Since this is a sequence learning task, we also provide it with the loadfactor $lf$ of each increment, where the loadfactor represents the percentage of the totally applied load. Once trained, the network predicts the non-local strain field $\bar{\varepsilon}^{NN}_{eq}$ and its derivative with respect to the local strain $\dfrac{\partial \bar{\varepsilon}^{NN}_{eq}}{\partial \varepsilon_{eq}}$. Here, it is crucial to emphasize that both the network inputs and outputs are quantities that are evaluated at the $Gauss \ points$ of the domain, and not at the nodes. The network outputs are then used to compute the element nodal contribution to the global Jacobian matrix $\bf{J}$ and residual vector $\bf{R}$. Therefore, $\bf{J}$ and $\bf{R}$ are informed by the non-local strain field, but their magnitude matches that of the unknown degrees-of-freedom vector, which is just the nodal displacements. Therefore, by computing the non-local strains but not treating them as nodal unknowns, it is possible to maintain the non-local character of the solution and also construct a smaller system of equations. The reduction in the number of nodal unknowns is the key that yields the computational savings of I-FENN since the resulting smaller equation system can be solved faster within an iterative numerical solver.  

The governing system of equations for I-FENN with non-local gradient damage is:

\begin{equation}
\begin{split}
    \bigl[ {\bf{J}} \bigl] \; \bigl[ {\delta \boldsymbol{u}} \bigl] = \bigl[ {\bf{R}} \bigl]
\end{split}
\label{IFENN_eqn_govern}
\end{equation}

\noindent where it is clearly shown that the unknown DoFs are just the nodal displacements. Numerical convergence at the $i^{th}$ iteration is achieved with a similar criterion as in the benchmark FEM solution (see Eqn. \ref{Eqn_convergence_criterion_FEM}):

\begin{equation}
    r_u = \frac{\prescript{}{i}{\| \delta{\boldsymbol{u}} \|_{2}}}{\prescript{}{1}{\| \delta{\boldsymbol{u}} \|_{2}}}  \leq tol
\label{Eqn_convergence_criterion_IFENN}
\end{equation}

The expression of the system residual vector $\bf{R}$ is:

\begin{equation}
    {\bf{R}} = \int_{\Omega} \left[{\bf{B}^u}\right]^{T} {\boldsymbol{\sigma}} \; d\Omega 
\label{IFENN_eqn_R}
\end{equation}

This residual is computed based on the internal stresses. In our numerical implementation we will be monitoring the reduction of its second norm as an additional check of the solver health, as follows:

\begin{equation}
    r_R = \dfrac{\prescript{}{i}{\| {\bf{R}} \|_{2}}}{\prescript{}{1}{\| {\bf{R}} \|_{2}}} 
\label{Eqn_convergence_criterion_IFENN_secondary}
\end{equation}

The system Jacobian matrix $\bf{J}$ can be calculated as: 

\begin{equation}
    \bf{J} =
    \left\{
        \begin{array}{ll}
        {\bf{K}} + \dfrac{\partial{\bf{K}}}{\partial{\boldsymbol{u}}}{\boldsymbol{u}}
        & \text {for full Newton-Raphson}  \\
        {\bf{K}} 
        & \text {for modified Newton-Raphson} 
        \end{array}
    \right.
\label{IFENN_eqn_J}
\end{equation}

\noindent where $\bf{K}$ is the global stiffness matrix and $\dfrac{\partial{\bf{K}}}{\partial{\boldsymbol{u}}}$ is the partial derivative of this matrix with respect to the nodal displacements. The expressions of these terms are:

\begin{equation}
    {\bf{K}} = \int_\Omega \! \left[{\bf{B}^u}\right]^{T} {(1-\boldsymbol{d}^{NN})} \, {\boldsymbol{C}} \, {\bf{B}^{u}} \, \mathrm{d}\Omega
\label{IFENN_eqn_K}
\end{equation}

\begin{equation}
    \dfrac{\partial{\bf{K}}}{\partial{\boldsymbol{u}}} = \int_\Omega \! \left[{\bf{B}^u}\right]^{T} {\boldsymbol{C}} \, \left(-\dfrac{\partial{\boldsymbol{d}^{NN}}}{\partial {\boldsymbol{u}}}\right) \, {\bf{B}^{u}} \, \mathrm{d}\Omega
\label{IFENN_eqn_pKpu}
\end{equation}

Here, we note that Eqns. \ref{IFENN_eqn_govern} - \ref{IFENN_eqn_pKpu} present a clear resemblance to the governing system of equations for the case of local damage \cite{lemaitrebook}, however, the non-local dispersion of the strain and damage fields is embedded in the numerical solution through the $d^{NN}$ and $\dfrac{\partial{d^{NN}}}{\partial {u}}$ terms. The damage variable is readily computed based on $\bar{\varepsilon}^{NN}_{eq}$ and the governing damage law, $d^{NN} = d(\bar{\varepsilon}^{NN}_{eq})$. To compute the partial derivative term, we utilize the chain rule as follows:

\begin{equation}
    \dfrac{\partial{d^{NN}}}{\partial u} = 
    \dfrac{\partial{d^{NN}}}{\partial{{\bar{\varepsilon}}^{NN}_{eq}}} \; 
    \dfrac{\partial{\bar{\varepsilon}^{NN}_{eq}}}{\partial{{\varepsilon}_{eq}}}       \; 
    \dfrac{\partial{\varepsilon_{eq}}}{\partial{\varepsilon_{ij}}}               \; 
    \dfrac{\partial{\varepsilon_{ij}}}{\partial u}
\label{FEM_Dddu_prop}
\end{equation}

\noindent where $\dfrac{\partial{d^{NN}}}{\partial{{\bar{\varepsilon}}^{NN}_{eq}}}$ is given by the damage law, $\dfrac{\partial{\bar{\varepsilon}^{NN}_{eq}}}{\partial{{\varepsilon}_{eq}}}$ is output by the neural network, $\dfrac{\partial{\varepsilon_{eq}}}{\partial{\varepsilon_{ij}}}$ depends on the local equivalent strain definition, and $\dfrac{\partial{\varepsilon_{ij}}}{\partial u}$ is equivalent to the shape function derivative matrix ${\bf{B}^{u}}$. 

Here, we emphasize that including the $\dfrac{\partial{\bf{K}}}{\partial{\boldsymbol{u}}}$ term in the $\bf{J}$ computation yields an approach equivalent to a full Newton-Raphson scheme. The user has the flexibility of omitting this term, arriving, therefore, at a modified Newton-Raphson version of the solution. Generally, the full Newton-Raphson is expected to converge in fewer iterations than its modified version \cite{crisfield1979faster}. However, if the neural network is not able to compute with sufficient accuracy the $\dfrac{\partial{\bar{\varepsilon}^{NN}_{eq}}}{\partial{{\varepsilon}_{eq}}}$ term, then the full NR may fail to converge in the first place. This is extensively discussed in Section \ref{Section:Results}, where we present several cases analyzed with both the full and the modified NR scheme.

\section{Temporal Convolutional Network (TCN) for I-FENN}
\label{Section:TCN}

In the presented setup, a sequence-to-sequence architecture is selected for the neural network within I-FENN, since our goal is to capture the history response of the domain. Here we adopt the Temporal Convolutional Network (TCN). Below we provide our reasoning behind this choice, we present the network architecture, we elaborate on the data-driven vs physics-informed formulation of its training, and we discuss input normalization and output un-normalization techniques.

\subsection{TCN relevance and architecture}
\label{Section:TCN_relevance_architecture}

First proposed by Bai et al. \cite{bai2018empirical}, TCNs are a special case of convolutional neural networks (CNNs) that were developed for sequence modeling. Let us suppose that based on an input sequence $x_{1},...,x_{T}$, one wishes to predict an output sequence $y_{1},...,y_{T}$ of the same length. The TCN design is based on the fundamental principle that the network prediction at load step $t$ is informed only by the current and past load increments, while it is unaware of any information stemming from the ``future''. This autoregressive character of the network is what makes it particularly suitable for non-linear problems in engineering mechanics, such as damage propagation, where capturing path-dependence requires knowledge of the past stages of the domain \cite{abueidda2023fenn, friemann2023micromechanics, mozaffar2019deep}. Also, as it is discussed in more detail in Section \ref{Section:IFENN_worfklow}, I-FENN invokes the trained TCN at every increment as the analysis proceeds, zero-padding the unseen increments in the input sequence. Technically, this implies a repeated and continuous change in the non-zero (past, current) and zero (future) entries of the input sequence. Therefore, we need to employ a network that can perform reasonably well under this condition, while also being able to make accurate predictions on the current increment. TCNs have this flexibility, which combined with their inherent ability to capture time-dependent causality, renders them a conceptually well-suited tool for our problem. 

\begin{figure}
    \centering
    \includegraphics[width=0.8\textwidth]{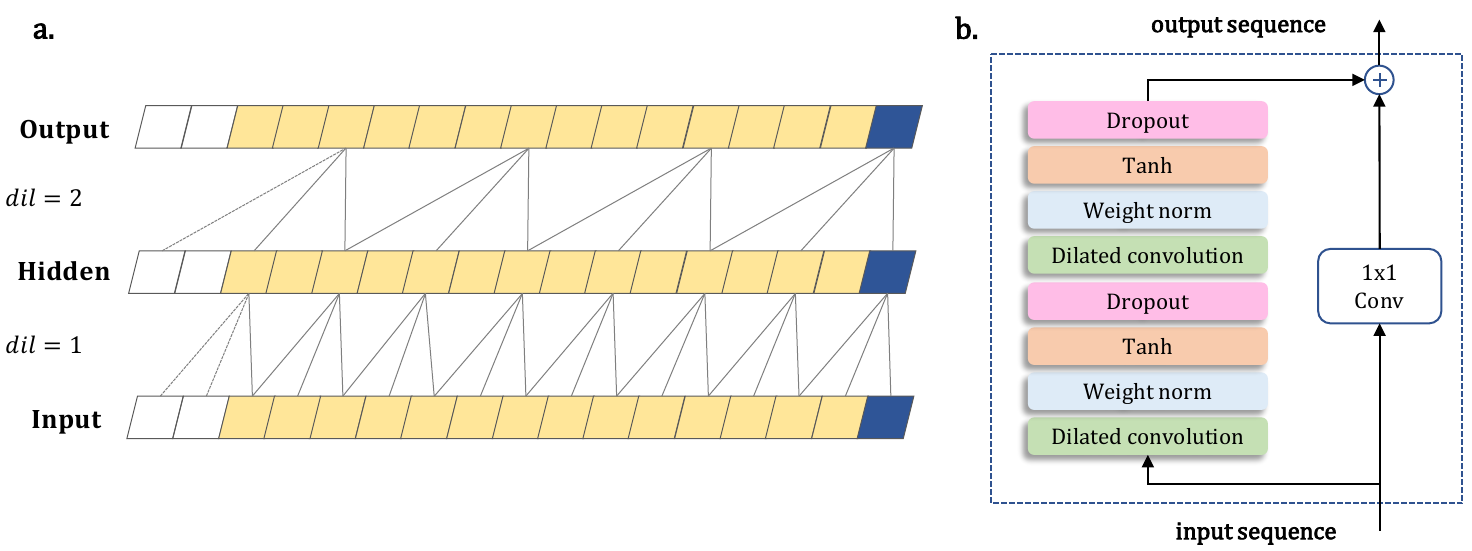}
    \caption{Architecture of the TCN: {\bf{a}}. Stack of dilated causal convolutions with dilation factors $dil = 1, 2$ and kernel size $ker = 3$. {\bf{b}}. TCN residual block with added layers of weight normalization, activation function and dropout in the main path, and an additional residual connection.}
    \label{Fig:Figure_TCN_architecture}
\end{figure}

The architecture of a sample TCN is depicted in Fig. \ref{Fig:Figure_TCN_architecture}. The basic functionality of the TCN is a 1D convolutional operation that is applied at each network layer. This operation is given as:

\begin{equation}
    F(s) = \sum_{i = 0}^{k_{size} - 1} k(i) \cdot {\bf{x}}_{s - dil \cdot i}
\label{TCN_Conv1}
\end{equation}

\noindent where $F$ is the output of the convolution on the element $s$ of the sequence $\bf{x}$, $k$ is the kernel (or filter), $k_{size}$ is the kernel size, $dil$ is the dilation factor, $i$ is the current position in the kernel, and $s - dil \cdot i$ represents the steps skipped from the current position which causes the dilation effect. Bai et al. \cite{bai2018empirical} also proposed an enhanced version of the main TCN block, which is shown in Fig \ref{Fig:Figure_TCN_architecture}b. In this version, which is adopted here, information flows through two pathways between the input and the output layers: a main path and a residual connection. The main path consists of a sequence of the following layers (repeated twice): the basic 1D dilated convolution \cite{yu2015multi}, weight normalization \cite{salimans2016weight}, activation function \cite{apicella2021survey} and dropout \cite{srivastava2013improving}. The latter three layers are rather standard practices in deep learning and generally aid the network performance. We note that here we set dropout equal to zero, and we choose the hyperbolic tangent ($tanh()$) for the non-linear activation. The latter choice is dictated by the presence of second-order partial derivatives in the cost function of the network. This requires an activation function that is at least twice differentiable, otherwise these derivatives trivially reduce to zero. Finally, the residual connection allows the information to skip the main path and be transferred directly to the output sequence. This feature, which has been shown to help prevent the vanishing gradients effect, consists of a 1x1 convolution on the input sequence and ensures avoiding a mismatch in the input and output data dimensionality. For more general information on the TCN architecture, the reader is referred to \cite{bai2018empirical}.

\subsection{Data-driven and physics-informed training}
\label{Section:TCN_training}

\subsubsection{Loss function}
\label{Section:TCN_loss}

The objective of the TCN is to approximate a target non-local strain field $\bar{\varepsilon}_{eq,true}$ with a predicted field $\bar{\varepsilon}_{eq}^{NN}$. This process entails adjusting the network weights $\mathcal{\phi}$ in order to minimize a loss function ${\boldsymbol{\mathcal{L}}}$:

\begin{equation}
    \mathcal{\phi}^{*} = \arg \min_{\mathcal{\phi}} {\boldsymbol{\mathcal{L}}}(\mathcal{\phi})
\label{TCN_argmin_params}
\end{equation}

In a $data-driven$ approach, the values of the $\bar{\varepsilon}_{eq,true}$ field are available. Denoting with $||.||_{2}$ the second norm of a vector, the data term of the loss function can be expressed as:

\begin{equation}
    {\boldsymbol{\mathcal{L}}}_{DATA} = || \bar{\varepsilon}_{eq}^{NN} - \bar{\varepsilon}_{eq,true} ||_{2}
\label{TCN_L_DATA}
\end{equation}

Over the last few years, and mainly fueled by the absence of large datasets in the engineering world, a paradigm shift has occurred in the way machine learning tools have been utilized in this field. Neural networks can be trained as surrogate models to approximate a physical relationship between the state variables of interest, even without any labeled data. In this case, the networks are labeled as $physics-informed$ \cite{cuomo2022scientific, cai2021physics}, and the training objective is the minimization of the residual of the governing PDE at the collocation points and its accompanying boundary condition at the boundary nodes of the domain. Physics-informed TCNs were first introduced in \cite{abueidda2023fenn} and showed superior performance to the simpler fully-connected architecture. For the problem considered here, the loss function can then be augmented with the following physics-based terms:

\begin{subequations}
    \begin{equation}
    {\boldsymbol{\mathcal{L}}}_{PDE} = \left \Vert 
    \bar{\varepsilon}_{eq}^{NN} - g \cdot \left(
    \dfrac{\partial^{2}{\bar{\varepsilon}}}{\partial{x}^{2}} +    \dfrac{\partial^{2}{\bar{\varepsilon}}}{\partial{y}^{2}} 
    \right) - \varepsilon_{eq} \right \Vert_{2}
    \label{TCN_L_PDE}
    \end{equation}

    \begin{equation}
    {\boldsymbol{\mathcal{L}}}_{BCs} = \left \Vert \dfrac{\partial\bar{\varepsilon}_{eq}^{NN}}{\partial{x}} \cdot {\bf{i}} + \dfrac{\partial\bar{\varepsilon}_{eq}^{NN}}{\partial{y}} \cdot {\bf{j}} \right \Vert_{2}
    \label{TCN_L_BCs}
    \end{equation}
\end{subequations}

\noindent where $\bf{i}$ and $\bf{j}$ are the outward unit vectors in the $x$ and $y$ directions. Thus, in the more general case where information from both the data and the physics is accounted for, the loss function can be expressed as:

\begin{equation}
    {\boldsymbol{\mathcal{L}}} = 
    w_{D} \cdot {\boldsymbol{\mathcal{L}}}_{DATA} + w_{P}  \cdot \left( {\boldsymbol{\mathcal{L}}}_{PDE} + {\boldsymbol{\mathcal{L}}}_{BCs} \right)
\label{TCN_L_TOTAL}
\end{equation}

\noindent where $w_{D}$ and $w_{P}$ are the weighting factors for the data and physics loss terms respectively. 

\subsubsection{Spatial gradients computation}
\label{Section:TCN_gradients}

Evidently, the cornerstone of the physics-informed training is the computation of the partial derivatives in the loss function, and automatic differentiation (AD) is the most commonly adopted method to compute these terms \cite{van2018automatic}. AD records the sequence of arithmetic operations in a forward pass, and uses the chain rule principle to compute the partial derivatives of the output variable with respect to any intermediate variable. Most platforms already use AD during the back-propagation stage in order to adjust the network weights \cite{baydin2018automatic, paszke2017automatic}, and therefore AD can be readily implemented to compute the spatial derivatives of the output variable as well. However, this method significantly amplifies the computational training time if higher-order partial derivatives need to be computed. 

As shown in the study of He at al \cite{he2023use}, an alternative way of computing the spatial derivatives is to discretize the computational domain with isoparametric elements, similar to standard FEM, and then use the element shape functions (SF) and Gauss quadrature to compute the derivatives. This approach has shown to be more stable than its AD counterpart \cite{he2023use}, but its major drawback is the tedious additional task of formulating the finite element discretization. However, this challenge is, by default, overcome in our case. This is simply because our TCN training domain has already been constructed with the finite element method, the collocation points in our TCN dataset correspond exactly to the FEM Gauss points, and the transformation from natural to physical coordinates is inherently available. This feature enables us to compute $a \ priori$ the spatial partial derivatives. The only key requirement we need to adhere to is the utilization of quadratic elements in the FEM discretization, which is necessary to compute non-zero higher-order derivatives at each Gauss point.

Since the implementation of AD is straightforward and its computational cost for the calculation of first-order derivatives is relatively small, we use AD to compute the first-order partial derivatives $\dfrac{\partial\bar{\varepsilon}}{\partial{x}}$, $\dfrac{\partial\bar{\varepsilon}}{\partial{y}}$, $\dfrac{\partial\bar{\varepsilon}}{\partial{\varepsilon}}$. The  Laplacian term in Eqn. \ref{TCN_L_PDE} is computed with the SF approach, since this is the main AD computational bottleneck. For our FEM mesh, we assume 8-node quadrilateral finite elements with 9 integration points. We introduce the following notation: $\xi$, $\eta$ are the natural coordinates of the isoparametric element, and $x$, $y$ are the physical coordinates of the Gauss point. We can show that the second order spatial derivatives of the non-local strain w.r.t. the physical coordinates at each Gauss point, $\dfrac{\partial^{2} \bar{\varepsilon}_{eq}}{\partial x^{2}}$, $\dfrac{\partial^{2} \bar{\varepsilon}_{eq}}{\partial y^{2}}$, and $\dfrac{\partial^{2} \bar{\varepsilon}_{eq}}{\partial x \partial y}$, can be obtained if one solves the following system:

\begin{equation}
\begin{bmatrix}
\dfrac{\partial^{2} \bar{\varepsilon}_{eq}}{\partial x^{2}} \\ \\
\dfrac{\partial^{2} \bar{\varepsilon}_{eq}}{\partial y^{2}} \\ \\
\dfrac{\partial^{2} \bar{\varepsilon}_{eq}}{\partial x \partial y} 
\end{bmatrix} 
= 
\left(\begin{bmatrix} 
\left( \dfrac{\partial x}{\partial \xi} \right)^{2}                         &
\left( \dfrac{\partial y}{\partial \xi} \right)^{2}                         & 
2 \dfrac{\partial x}{\partial \xi} \dfrac{\partial y}{\partial \xi}      \\ & \\
\left( \dfrac{\partial x}{\partial \eta} \right)^{2}                        & 
\left( \dfrac{\partial y}{\partial \eta} \right)^{2}                        &
2 \dfrac{\partial x}{\partial \eta} \dfrac{\partial y}{\partial \eta}    \\ & \\
\dfrac{\partial x}{\partial \xi} \dfrac{\partial x}{\partial \eta}          & 
\dfrac{\partial y}{\partial \xi} \dfrac{\partial y}{\partial \eta}          & 
\dfrac{\partial x}{\partial \xi} \dfrac{\partial y}{\partial \eta} + \dfrac{\partial x}{\partial \eta} \dfrac{\partial y}{\partial \xi} 
\end{bmatrix}\right)^{-1} \cdot
\begin{bmatrix} 
\dfrac{\partial^{2} \bar{\varepsilon}_{eq}}{\partial \xi^{2}} - 
\dfrac{\partial \bar{\varepsilon}_{eq}}{\partial x}\dfrac{\partial^{2} x}{\partial \xi^{2}} - 
\dfrac{\partial \bar{\varepsilon}_{eq}}{\partial y}\dfrac{\partial^{2} y}{\partial \xi^{2}} \\ \\ 
\dfrac{\partial^{2} \bar{\varepsilon}_{eq}}{\partial \eta^{2}} - 
\dfrac{\partial \bar{\varepsilon}_{eq}}{\partial x}\dfrac{\partial^{2} x}{\partial \eta^{2}} - 
\dfrac{\partial \bar{\varepsilon}_{eq}}{\partial y}\dfrac{\partial^{2} y}{\partial \eta^{2}} \\ \\
\dfrac{\partial^{2} \bar{\varepsilon}_{eq}}{\partial \xi\eta} - 
\dfrac{\partial \bar{\varepsilon}_{eq}}{\partial x}\dfrac{\partial^{2} x}{\partial \xi \partial \eta} - 
\dfrac{\partial \bar{\varepsilon}_{eq}}{\partial y}\dfrac{\partial^{2} y}{\partial \xi \partial \eta}
\end{bmatrix}
\label{Eqn:SF_derivation}
\end{equation}

The complete derivation of Eqn. \ref{Eqn:SF_derivation} as well as the mathematical formulas of all the terms involved can be found in \ref{Appendix:SF_derivation}.

\subsection{Normalization and un-normalization of input and output variables}
\label{Section:DataNorm}

The objective of input data normalization is to bridge the gap between the potentially different scales in the input variables. The goal of this technique is to expedite the network training and improve its performance, and it has become a standard practice in many data-driven problems \cite{singh2020investigating}. However, in physics-informed problems such as the one considered here, normalization of the input variables has a straightforward impact on the cost function definition because the normalized input values are directly embedded in the physics-based loss terms. Therefore, satisfying the governing physical relationship while accounting for normalized input will evidently yield scaled output. These need to be brought back to the original scale through an inverse process in order to assess the actual predictions. This is a particularly critical step for I-FENN because the TCN outputs (predictions and partial derivatives) are used in the FEM analysis for the Jacobian and residual computations, and therefore their correct integration is imperative. 

Let us denote with ${\boldsymbol{T}}(v)$ a general normalization function on the variable $v$. In our problem, the values of the strain field $\varepsilon_{eq}$ can be several orders of magnitude smaller than the coordinates $x, y$. Therefore we choose to scale $\varepsilon_{eq}$ accordingly, while the coordinates remain untouched. Based on the governing Eqn. \ref{Eqn_NonlocalGradientPDE_1}, we can then write:

\begin{equation}
\begin{split}
    {\boldsymbol{T}} \left( \bar{\varepsilon}_{eq} - g \left( \frac{\partial^{2}{\bar{\varepsilon}_{eq}}}{\partial{x}^{2}} + \frac{\partial^{2}{\bar{\varepsilon}_{eq}}}{\partial{y}^{2}}\right) \right) = {\boldsymbol{T}} \left( \varepsilon_{eq} \right) 
\label{nonlocalGradientPDE_lineartransf1}
\end{split}
\end{equation}

\noindent which implies that if a normalization ${\boldsymbol{T}}(.)$ is performed on the local strain field, then the same operation needs to be applied to the diffused part of the PDE in order to satisfy the physical law. In principle ${\boldsymbol{T}}$ can be any linear or non-linear function. Here we constrain our investigation to linear functions only, and ${\boldsymbol{T}}(.)$ is given as ${\boldsymbol{T}}(v) = a \cdot (v) + b$. Expanding Eqn. \ref{nonlocalGradientPDE_lineartransf1} and denoting the scaled quantities with the superscript $(.)^{'}$, we can then write:


\begin{equation}
\begin{split}
    a \left( \bar{\varepsilon}_{eq} - g \left( \frac{\partial^{2}{\bar{\varepsilon}_{eq}}}{\partial{x}^{2}} + \frac{\partial^{2}{\bar{\varepsilon}_{eq}}}{\partial{y}^{2}}\right) \right) + b = a \varepsilon_{eq} + b \Rightarrow \underbrace{\left( a \bar{\varepsilon}_{eq} + b \right)}_{{\bar{\varepsilon}_{eq}^{'}}} - a g \left( \frac{\partial^{2}{\bar{\varepsilon}_{eq}}}{\partial{x}^{2}} + \frac{\partial^{2}{\bar{\varepsilon}_{eq}}}{\partial{y}^{2}}\right) = \underbrace{a \varepsilon_{eq} + b}_{{\varepsilon_{eq}^{'}}}
\label{Minmax_transf3}
\end{split}
\end{equation}

Equation \ref{Minmax_transf3} is the one used for the TCN training, which takes as input the scaled local strain $\varepsilon_{eq}^{'}$ and predicts a scaled non-local strain $\bar{\varepsilon_{eq}}^{'}$. We underline that the Laplacian term $\left( \dfrac{\partial^{2}{\bar{\varepsilon}_{eq}}}{\partial{x}^{2}} + \dfrac{\partial^{2}{\bar{\varepsilon}_{eq}}}{\partial{y}^{2}}\right)$ has already been computed using the SF approach. The unscaled predictions $\bar{\varepsilon}_{eq}$ and the corresponding partial derivatives can be then computed through the inverse operation $\boldsymbol{T}(.)^{-1}$ and the chain rule respectively as follows:

\begin{equation}
\begin{split}
    \bar{\varepsilon}_{eq} = a^{-1} \left( \bar{\varepsilon}_{eq}^{'} - b \right)
\label{Minmax_transf8}
\end{split}
\end{equation}

\begin{equation}
\begin{split}
    \frac{\partial{\bar{\varepsilon}_{eq}}}{\partial{\varepsilon_{eq}}} = \frac{\partial{\bar{\varepsilon}_{eq}}}{\partial\bar{\varepsilon}_{eq}^{'}} \; \; \frac{\partial{\bar{\varepsilon}_{eq}^{'}}}{\partial{\varepsilon_{eq}^{'}}} \; \; \frac{\partial{\varepsilon}_{eq}^{'}}{\partial{\varepsilon_{eq}}} = a^{-1} \; \; \frac{\partial{\bar{\varepsilon}_{eq}^{'}}}{\partial{\varepsilon_{eq}^{'}}} \; \; 
    a = \frac{\partial{\bar{\varepsilon}_{eq}^{'}}}{\partial{\varepsilon_{eq}^{'}}} 
\end{split}
\end{equation}

\begin{equation}
\begin{split}
    \frac{\partial{\bar{\varepsilon}_{eq}}}{\partial{x}} = \frac{\partial{\bar{\varepsilon}_{eq}}}{\partial\bar{\varepsilon}_{eq}^{'}} \; \; \frac{\partial\bar{\varepsilon}_{eq}^{'}}{\partial{x}} = a^{-1} \; \; \frac{\partial\bar{\varepsilon}_{eq}^{'}}{\partial{x}} \; \; , \; \; \mathrm{similar \ for} \; \;  \frac{\partial{\bar{\varepsilon}_{eq}}}{\partial{y}}
\end{split}
\end{equation}

\noindent where we note that $\dfrac{\partial{\bar{\varepsilon}_{eq}^{'}}}{\partial{\varepsilon_{eq}^{'}}}$, $\dfrac{\partial{\bar{\varepsilon}_{eq}^{'}}}{\partial{x}}$, $\dfrac{\partial{\bar{\varepsilon}_{eq}^{'}}}{\partial{y}}$ are computed with automatic differentiation. 

In this study we consider three cases of linear normalization functions:

\begin{itemize}
    
    \item Constant decimal (CD) scaling: 
    
    ${\varepsilon}_{eq}^{'} = {\boldsymbol{T}}({\varepsilon}_{eq}) = 10^{k} \cdot {\varepsilon}_{eq}$. The coefficients of the normalization function are $a = 10^{k}$ and $b = 0$, and the local strain field of the entire loading history is uniformly multiplied with the constant decimal scaling factor $10^k$. This normalization was also adopted in our previous work \cite{pantidis2023integrated, pantidis2023116160}.
    
    \item Varying multiplication (VM) scaling: 
    
    ${\varepsilon}_{eq}^{'} = {\boldsymbol{T}}({\varepsilon}_{eq}) = p \cdot {\varepsilon}_{eq}$. In this case $a = p$ and $b = 0$, where $p$ is a varying multiplication factor with a different value at each load increment. The evolution of $p$ is presented along with our numerical implementation in Section \ref{Section:Results}. 

    \item Min-max (MM) scaling between $[0-new\_max]$: 
    
    ${\varepsilon}_{eq}^{'} = {\boldsymbol{T}}({\varepsilon}_{eq}) = \dfrac{new\_max}{max({\varepsilon}_{eq}) - min({\varepsilon}_{eq})} \cdot {\varepsilon}_{eq} + \left( - \dfrac{new\_max \cdot min({\varepsilon}_{eq})}{max({\varepsilon}_{eq}) - min({\varepsilon}_{eq})}\right)$. In this case ${\varepsilon}_{eq}$ at each increment is normalized based on the minimum and maximum value at that load increment. 
    
\end{itemize}

\subsection{Hyper-parameter optimization}
\label{Section:HyperOpt}

The performance of a neural network is largely dictated by the choice of its hyperparameters, such as dimensions and number of layers \cite{pantidis2023116160}, learning rate and training algorithms \cite{abdolrasol2021artificial}, activation functions \cite{apicella2021survey} and weight initialization \cite{kumar2017weight}. Manual tuning and grid search are a straightforward way to explore this search space, but these methods are cumbersome and computationally inefficient \cite{bergstra2011algorithms}. More sophisticated approaches such as Neural Architecture Search \cite{elsken2019neural} or Bayesian optimization \cite{snoek2012practical} can be implemented, but the evident drawback of these methods is the significant additional expense that is imposed on the training process.

In this study we keep the TCN hyperparameters fixed for the vast majority of our analysis, since the focal point is the functionality of the TCN-based I-FENN. However, we also extend our code to be compatible with a commercially available optimization package, in an effort to further increase the flexibility of the framework's end-user. For this work we select $Hyperopt$ \cite{bergstra2013making}, which enables the simultaneous tuning of multiple hyper-parameters by employing an advanced search algorithm. In our study we adopt the Tree of Parzen Estimators (TPE) algorithm, an iterative process that creates a probabilistic model using the history of the evaluated hyperparameters and then suggests the next set of hyperparameters to assess. Using this approach, we present a sample case in Section \ref{Section:SNS}, where we seek to optimize the dimension-related parameters of the TCN.

\section{I-FENN workflow with TCNs for non-local gradient damage}
\label{Section:IFENN_worfklow}

So far we have presented the relevant tools and practices, and we have introduced the corresponding mathematical notation. In this section we assemble these components and we present a detailed description of the I-FENN workflow. 

\subsection{Data Generation}

\begin{figure}[b]
    \centering
    \includegraphics[width=0.95\textwidth]{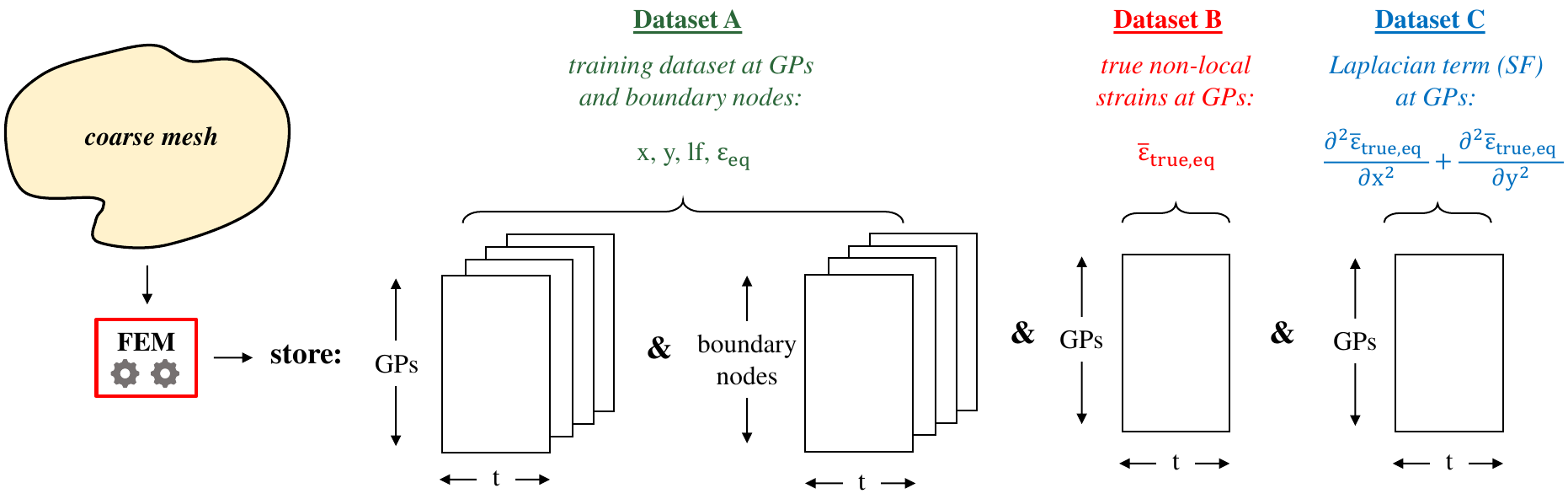}
    \caption{Schematic overview of the Data Generation stage of I-FENN.}
    \label{Fig:Figure_IFENN_Stage1}
\end{figure} 

The first step is to perform an FEM analysis on a coarse idealization of the investigated domain. The outcomes of this analysis are: a) a 3D tensor that contains the coordinates, local strains and loadfactor values for the Gauss points and boundary nodes for all the load increments, b) the true non-local strain field at the Gauss points, and c) the Laplacian term values at the Gauss points. These are denoted as Datasets A (green), B (red) and C (blue) respectively in Fig. \ref{Fig:Figure_IFENN_Stage1}. Here we make the following remarks. First, we note that non-temporal values (such as the coordinates) remain constant during the analysis, but we still store them in the input sequence tensor in order to facilitate the TCN training. Also, while Dataset A is essential for all models, Datasets B and C are optional. Dataset B is required only if the user opts to use a data-driven training, otherwise it can be omitted. Dataset C is needed only if the user chooses both to add the physics in the loss definition (see Eqn. \ref{TCN_L_TOTAL}) and use the shape function approach for these derivatives. In that case we underline that higher-order elements are required for this preliminary analysis, and in this paper we use quadratic elements. In any other case, which is either only data-driven training or physics-informed with just AD, then first-order elements suffice.

\subsection{Network Training}

\begin{figure}[H]
    \centering
    \includegraphics[width=0.95\textwidth]{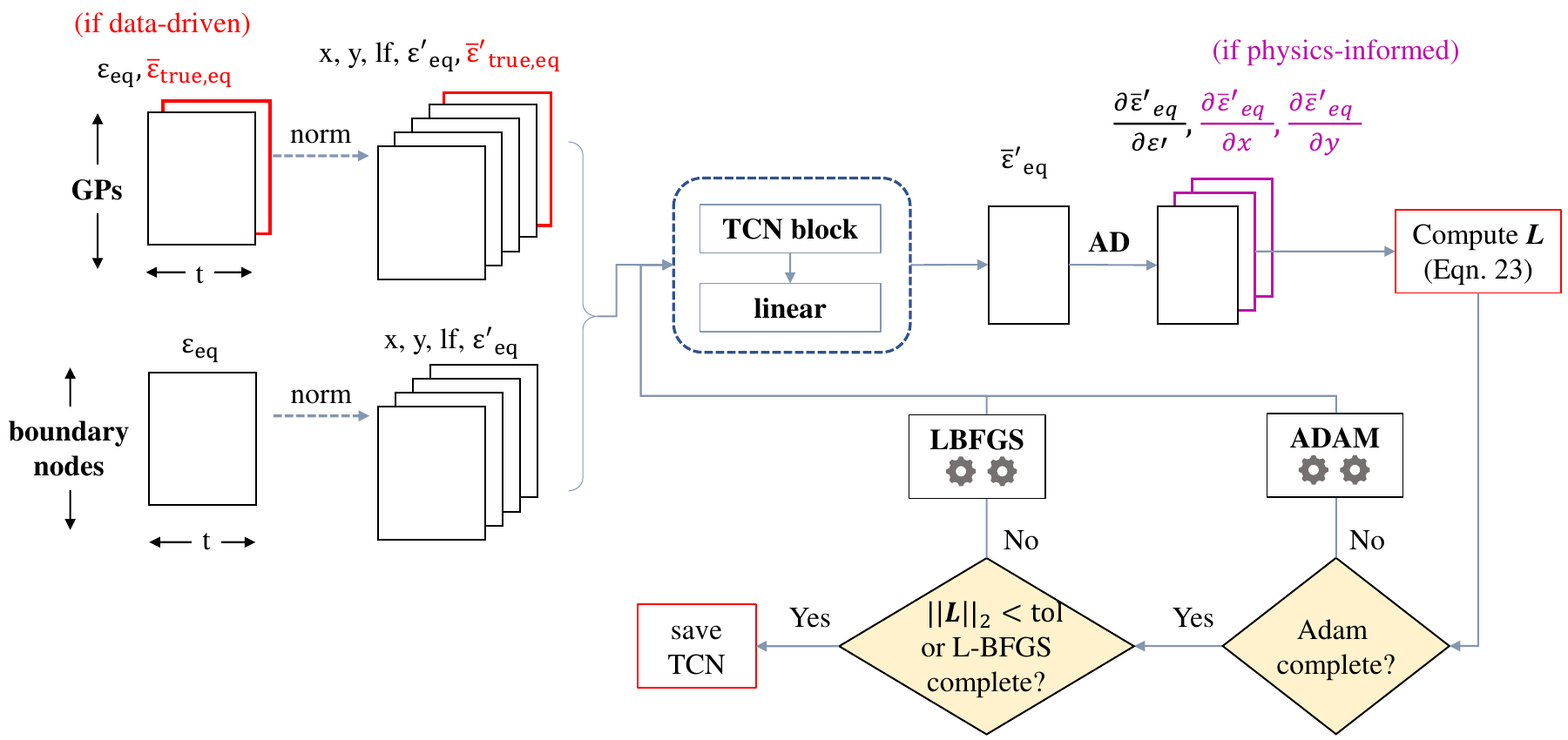}
    \caption{Schematic overview of the Network Training stage of I-FENN.}
    \label{Fig:Figure_IFENN_Stage2}
\end{figure}

The second step is the training of the TCN network. This is the stage where the user navigates through the training-related options that were presented in the previous section. In this paper, we adopt the scheme that is presented in Fig. \ref{Fig:Figure_IFENN_Stage2}. Once the training dataset is created, we first normalize the local strain field at the Gauss points and boundary nodes, as well as the true non-local strains at the Gauss points (only for data-driven training). The training dataset is then passed into one TCN residual block with an architecture as shown in Fig. \ref{Fig:Figure_TCN_architecture}, and the TCN output is then fed into a linear (fully-connected) layer. The output of this forward pass is the predicted non-local strain field at the Gauss points and the boundary nodes. Using automatic differentiation, we compute the first-order partial derivatives of the non-local strain with respect to the local strain $\dfrac{\partial{\bar{\varepsilon}_{eq}}}{\partial{\varepsilon_{eq}}}$ for the Gauss points, and with respect to the coordinates $\dfrac{\partial{\bar{\varepsilon}_{eq}}}{\partial{x}}$, $\dfrac{\partial{\bar{\varepsilon}_{eq}}}{\partial{y}}$ for the boundary nodes. We underline that the latter two terms are needed only when we account for the physics-related boundary condition term in the loss function (Eqn. \ref{TCN_L_TOTAL}). We then compute the network loss and use two commonly adopted optimizers to update the network parameters: we first utilize Adam \cite{kingma2014adam} for a predefined number of training epochs, and we then implement L-BFGS \cite{liu1989limited} until the loss function has converged below the algorithm tolerance. Once the training is complete, we store the trained network, which will be used in the next stage of IFENN.

\subsection{FEM analysis}

\begin{figure}[H]
    \centering
    \includegraphics[width=1\textwidth]{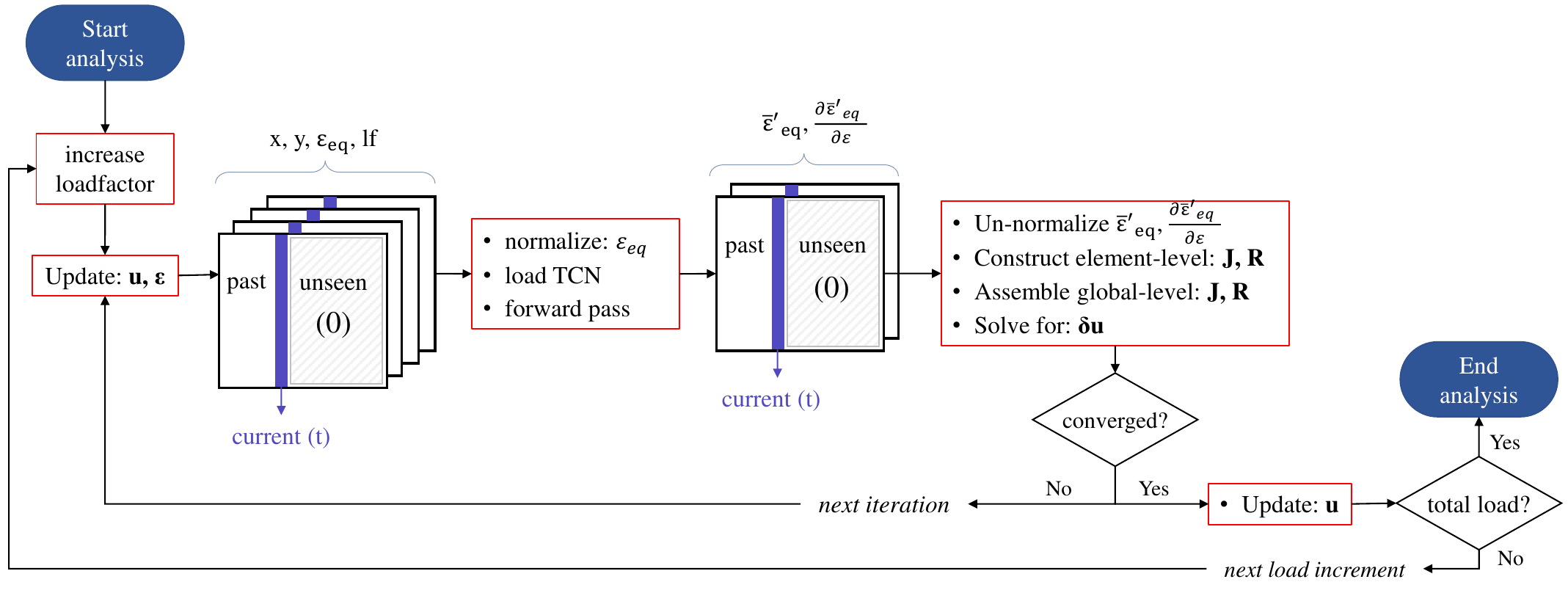}
    \caption{Schematic overview of the FEM analysis stage of I-FENN.}
    \label{Fig:Figure_IFENN_Stage3}
\end{figure}

The third step is the integration of the TCN within the FEM algorithm and the numerical simulation of the test model. A schematic view of this step is shown in Fig. \ref{Fig:Figure_IFENN_Stage3}. The same domain as in the first step is analyzed, which in this step is discretized using a finer mesh resolution. We note that either first- or higher-order elements can be used at this stage, regardless of the choice of elements in the training model. The analysis begins following the principles outlined in Section \ref{Section:IFENN_framework}. At the first iteration of any given increment, we update the prescribed displacements, and we compute the local strain field at the Gauss points. Then, the trained TCN is invoked and it is being fed with a tensor that contains the GP coordinates, local strains and loadfactor values until \textit{that} increment. In order to meet the requirements of size and dimensionality of the trained TCN, the remaining (future) increments in the input sequence are padded with zeros. The TCN performs a forward pass and computes the non-local strains and their derivatives w.r.t. the local at every integration point. The outputs that correspond to the current increment are accessed and un-normalized accordingly. These quantities are then used to construct the element-level Jacobian and residual, which are assembled into the global quantities. We then compute the incremental change of the nodal displacements and check their convergence. This process is repeated until this increment has converged, and the analysis can proceed to the next increment.

\section{Numerical results}
\label{Section:Results}

In this section, we present the results of the I-FENN implementation on three benchmark examples. Our main investigation model is a square domain with a single notch under tension loading. This model is used for the detailed exploration of our search space: different normalization techniques, data-driven vs physics-informed TCN training, I-FENN analysis with full and modified Newton-Raphson, computational savings against finer mesh resolutions. The other two models are a double-notch domain under tension and a single-notch domain under shear. In these cases, we capitalize on the lessons learned from the investigation model regarding the TCN training, and we focus our attention on the I-FENN analysis and computational gains. 

\subsection{Investigation Model: Single notch under tension (SNT)}
\label{Section:SNT}

\subsubsection{Training and testing models}
\label{Section:SNT_model}

\begin{figure}[H]
	\centering
	\includegraphics[width=1\textwidth]{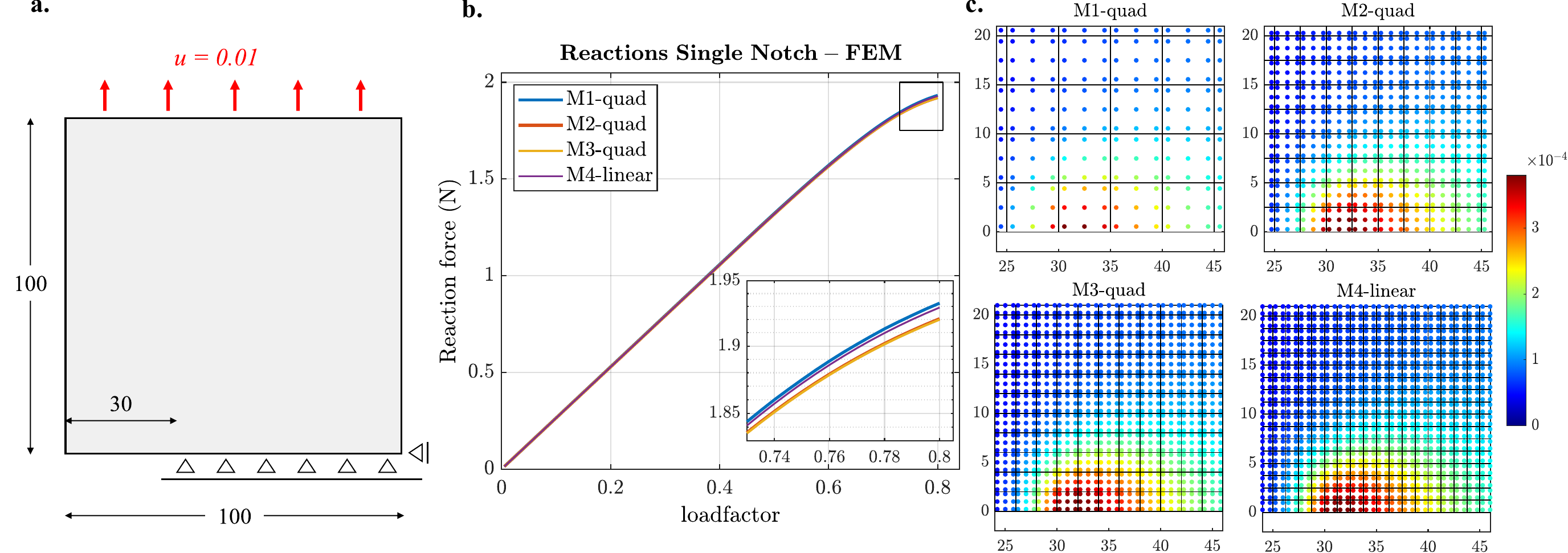}
	\caption{{\bf{a.}} Geometry, loading details, and boundary conditions for the single notch tension problem. Displacement units are in mm. {\bf{b.}} Reaction-loadfactor curves for all seven idealizations (3 cases with quadratic elements, 4 cases with linear elements). Force is measured in N. {\bf{c.}} Non-local strain profiles at the last increment of 4 representative models. We underline that the TCN will be trained only on the M1-quad dataset, and it will be used to predict the response of the finer and unseen mesh idealizations.}
	\label{Fig:Figure_SNT_geom_reactions_nonlocalstrains}
\end{figure}

The geometric and loading details of the single notch tension (SNT) problem are shown in Fig \ref{Fig:Figure_SNT_geom_reactions_nonlocalstrains}a. The domain has a square shape of size $100mm$x$100mm$, with the bottom edge unsupported between $x < 30mm$ and supported by rollers between $30mm < x < 100mm$. The top edge is subject to a tensile displacement of $u_{D} = 0.01mm$, incremented quasi-statically with constant load increments of $\Delta u_{D} = 0.005$. In all models discussed below, the load history is truncated at $u_{D} = 0.008mm$, which corresponds to a loadfactor $lf = 0.008/0.01 = 0.8$. The reason is that beyond that point the excessive non-linearities disrupted the constant load incrementation, which is a prerequisite for better utilization of the TCN.

\renewcommand{\arraystretch}{1}
\begin{table}[H]
\caption{Mesh details and utilization of training and testing models for the single notch tension problem.}
\label{Table:SNT_geometries} \centering
\begin{tabular}{c c c c c c}
    \hline
    {\bf{Model name}} & {\bf{Element order}} & {\bf{\# Nodes}} & {\bf{\# Elems.}} & {\bf{\# GPs}} & {\bf{Purpose}} \\   
    \hline
    M1-quad    & 2 & 1281 & 400 & 3600 & Training data \& TCN variations \\
    \hline
    M2-linear  & 1 & 1681 & 1600 & 6400 & I-FENN solver variations \\
    \hline
    M2-quad    & 2 & 4961 & 1600 & 14400 & Computational savings \\
    M3-quad    & 2 & 7701 & 2500 & 22500 & Computational savings \\
    M4-linear  & 1 & 6561 & 6400 & 25600 & Computational savings \\
    \hline
\end{tabular}
\end{table} 

The domain is discretized using a structured finite element mesh, with either linear or quadratic square elements and a constant element size $l_{elem}$. Four element sizes are selected: $l_{elem} = \{5.0, 2.5, 2.0, 1.25\} \ mm$, and the corresponding models are termed M1, M2, M3, and M4. Table \ref{Table:SNT_geometries} presents the mesh details for the generated models, which are utilized as follows:

\begin{itemize}
    \item M1-quad: Generate the TCN training dataset, and perform checks to a) justify why employing the normalization schemes is essential, and b) understand the impact of different loss function formulations and normalization schemes on the network accuracy.
    \item M2-linear: Explore the impact of different TCN variations on the IFENN performance, tested with both full and modified Newton-Raphson.
    \item M4-linear, M2-quad, M3-quad: Using one of the trained TCNs, compare their benchmark FEM solution against I-FENN and explore the computational savings.
\end{itemize}

In all models, the characteristic length is $l_{c} = 4mm$ and the strain threshold value is $\bar{\varepsilon}_{D} = 0.0001$. The Mazar's damage model \cite{mazars1986description} is used, with parameters $\alpha = 0.7$ and $\beta = 10^{4}$. For the SNT problem the equivalent strain is defined as $\varepsilon_{eq} = \sqrt{\langle\varepsilon_{1}\rangle^{2} + \langle\varepsilon_{2}\rangle^{2} + \langle\varepsilon_{3}\rangle^{2}}$, where $\varepsilon_{i, \ i = 1, 2, 3}$ are the principal strains and $\langle \, \cdot \: \rangle$ are the Macaulay brackets.

The benchmark FEM analysis of these models was performed and the reaction-loadfactor curves are shown in Fig. \ref{Fig:Figure_SNT_geom_reactions_nonlocalstrains}b. Their overlap is an evident demonstration of the mesh-independent character of the problem. Here, it is also worth visualizing the non-local strain contours at the end of the analysis. This step is carried out in Fig. \ref{Fig:Figure_SNT_geom_reactions_nonlocalstrains}c for few representative models: M1-quad, M2-quad, M3-quad and M4-linear. For better clarity, only a zoomed portion of the domain is shown ($25mm < x < 45mm$, $0mm < y < 20mm$), where the black grid indicates the edges of the elements and the markers are located at the Gauss points positions. The first observation is that the non-local strain follows the same diffused profile across all models, further proving the correctness of the numerical solution. Even more importantly, this graph shows the striking difference between our very coarse training dataset (M1-quad) and the significantly finer mesh resolutions that are used in the I-FENN implementation stage (such as M3-quad and M4-linear).

\subsubsection{Variations of the TCN training setup}
\label{Section:SNT_TCN_training}

\begin{figure}[H]
	\centering
	\includegraphics[width=1\textwidth]{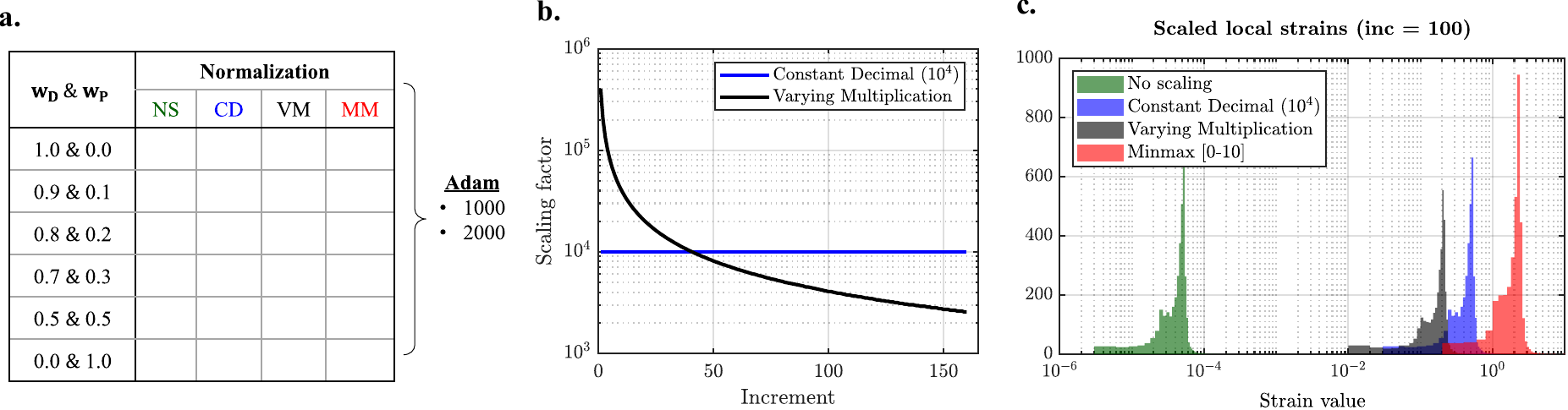}
	\caption{{\bf{a.}} Search space of the TCN setup. {\bf{b.}} Constant decimal and varying scaling factors during the load history. {\bf{c.}} Histograms of the unscaled and scaled local strain field at the $100^{th}$ load increment for the different normalization methods.}
	\label{Fig:Figure_TCN_variations}
\end{figure}

In this section we explore the impact of three main aspects of the TCN training: a) data-driven vs physics-informed formulation of the loss function, b) different input normalization and output un-normalization methods, and c) different values of Adam epochs. The search space is shown in Fig. \ref{Fig:Figure_TCN_variations}a. For the data-vs-physics aspect, we investigate 6 cases: a purely data-driven case (${w_{D}} = 1.0$ and ${w_{P}} = 0.0$), a purely physics-informed case (${w_{D}} = 0.0$ and ${w_{P}} = 1.0$), and four intermediate combinations. For the normalization aspect, we explore four scenarios: a) no scaling, b) constant decimal (CD) scaling with a $10^4$ factor, c) varying multiplication (VM) scaling where the coefficient $p$ is shown in Fig. \ref{Fig:Figure_TCN_variations}b, and d) min-max (MM) normalization between $[0-10]$. A sample example of local strain scaling is shown at Fig. \ref{Fig:Figure_TCN_variations}c for the $100^{th}$ increment of the load history, which helps visualizing the actual order of magnitude of the unscaled strain field as well as the data-shift for the other three methods. For the Adam epochs aspect, we investigate two cases: 1000 and 2000 training epochs, and we then allow the L-BFGS algorithm to run until convergence. We explore the search space in a full-grid approach, which yields 6 x 4 x 2 = 48 TCNs in total. In all these cases the network size stays the same: one hidden layer, $dil = 3$, $k_{size}= 12$, $num_{filters} = 6$.

In order to measure the predictive accuracy of the trained TCNs we use the relative squared error of the predictions, which is measured at each load increment and it is defined as follows:

\begin{equation}
    \bar{\varepsilon}_{RSE} = \sqrt{\sum_{1}^{No. GPs} \dfrac{\left( \bar{\varepsilon}_{i}^{NN} - \bar{\varepsilon}_{true,i} \right)^{2}}{\left( \bar{\varepsilon}_{true,i} \right)^2}}
\end{equation}

\begin{figure}[t]
	\centering
	\includegraphics[width=0.88\textwidth]{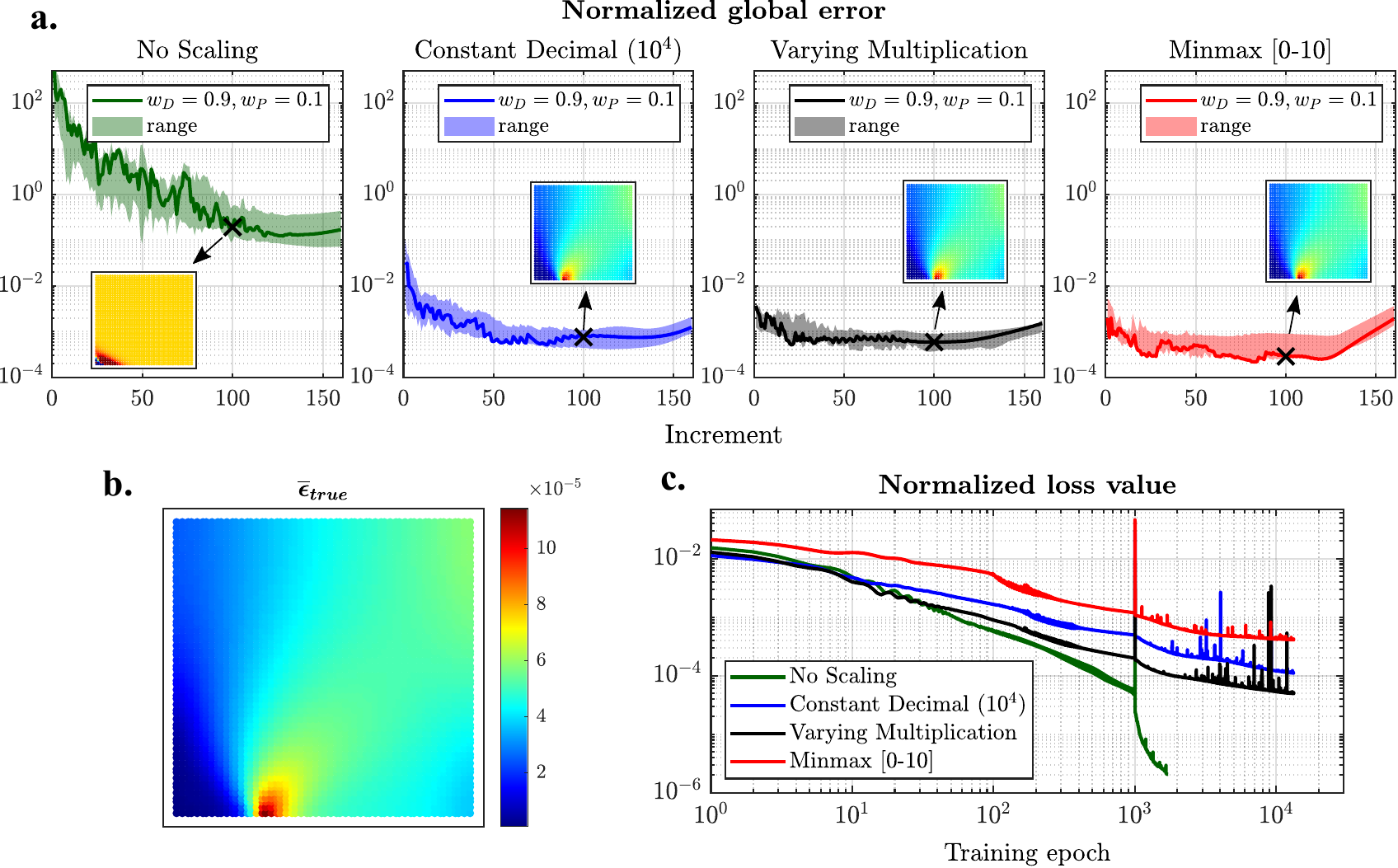}
	\caption{{\bf{a.}} Evolution of the TCN global error for different scaling methods. The shaded area is bounded by all the idealizations of each scaling. The solid line represents the model with 90\% data, 10\% physics and Adam = 1000 epochs. Inset plots show the predictions at the $100^{th}$ load increment. These plots show that the normalization/un-normalization step is a crucial prerequisite for successful training {\bf{b.}} True non-local strain profile at the $100^{th}$ increment. {\bf{c.}} Evolution of the TCN loss value for the selected solid-line models. This graph shows that caution must be exercised when evaluating this error metric for different scaling methods.}
	\label{Fig:Figure_Justifying_Scaling}
\end{figure}

In Fig. \ref{Fig:Figure_Justifying_Scaling}a we plot $\bar{\varepsilon}_{RSE}$ against the loading history, using the same dataset for predictions as for training (M1-quad). The shaded area denotes the patched region that is constructed from the 6 x 2 = 12 models of each scaling method. For illustration purposes, we also show the evolution of $\bar{\varepsilon}_{RSE}$ for the model trained with $90\%$ data, $10\%$ physics, and 1000 Adam epochs. One can immediately observe that the error in the unscaled case is several orders of magnitude higher than in any other case where scaling has been applied. This is a clear sign that incorporating these normalization and un-normalization schemes is an essential step for utilizing the TCN architecture, and it has substantially improved the predictive performance of the networks. This is even more clearly illustrated if we visualize the resulting predictions. For the sample case of the $100^{th}$ load increment, we plot the predicted non-local strains in the embedded contour plots of Fig. \ref{Fig:Figure_Justifying_Scaling}a. For comparison, we show the true values of the non-local strain and the corresponding color bar in Fig. \ref{Fig:Figure_Justifying_Scaling}b. We observe that the no-scaling approach has led to complete failure in the predictions, whereas in any of the scaling schemes, the predictions show an excellent agreement with the correct solution. Therefore, these results clearly demonstrate that including the normalization/un-normalization procedure is an unavoidable and crucial step for training the TCN successfully. 

Here we also need to make a note on the networks' loss behavior. In Fig. \ref{Fig:Figure_Justifying_Scaling}c we plot the loss history of the selected configuration ($w_{D} = 0.9$, $w_{P} = 0.1$, Adam = 1000) with the four different scalings. We see that it can be misleading to conclude on the training success based just on the loss reduction, and this is explained as follows. In the no-scaling case, the TCN receives input numbers in the order of $10^{-8}$ - $10^{-4}$, and it attempts to match them by predicting that small numbers as well. Therefore, their mismatch will be a low number, hence the low values of $\mathcal{L}$. This process however does not yield meaningful predictions, as it was evidently shown before. Larger numerical values in the input and output sequences imply a better chance of successful TCN training, and together with the un-scaling step $after$ the training is done, helps to restore accurately the correct numerical values of the non-local strains.

\begin{figure}[H]
	\centering
	\includegraphics[width=1\textwidth]{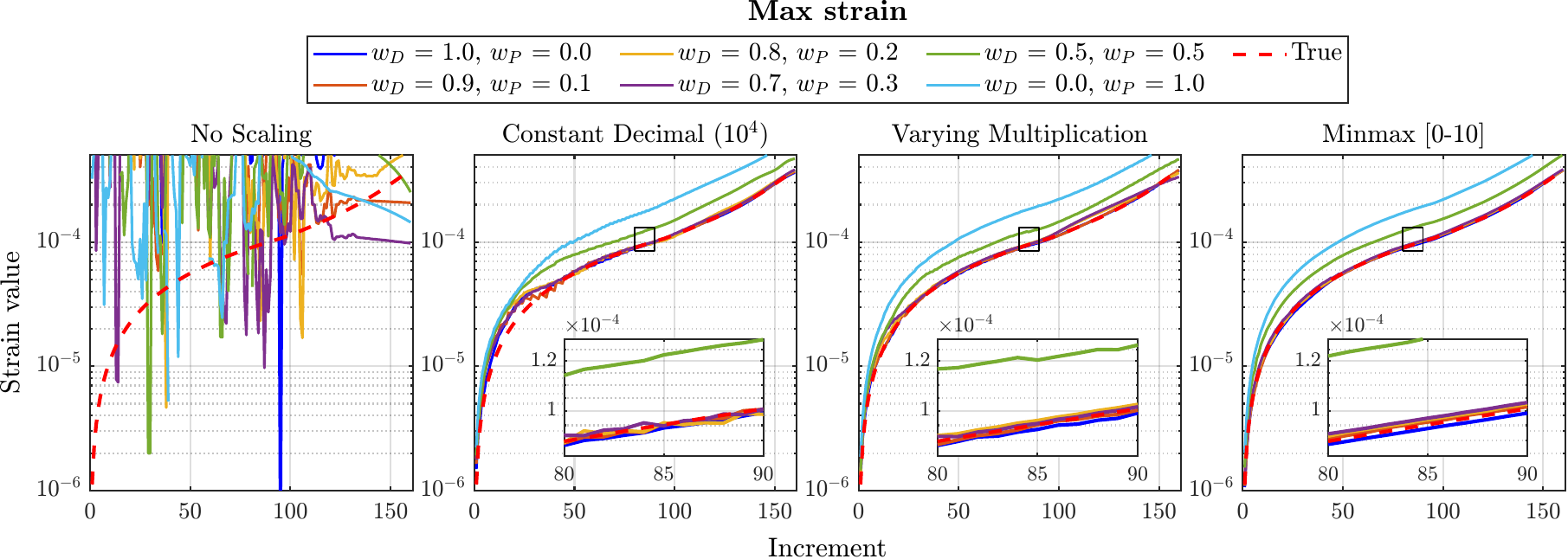}
	\caption{Evolution of the maximum non-local strain throughout the load history, for the different scaling methods and for all data-vs-physics combinations.}
	\label{Fig:Figure_Choosing_Scaling}
\end{figure}

We now perform one more check, in order to examine whether any of the scaling approaches or data-vs-physics combinations is more well-suited. A peculiarity of our damage mechanics problem is that damage is mainly driven by the maximum value of the non-local strain, and it is therefore very important to capture that value with sufficient accuracy \cite{pantidis2023116160}. In Fig. \ref{Fig:Figure_Choosing_Scaling} we plot $\bar{\varepsilon}_{max}$ for all the trained TCNs, where the sub-plots correspond to the four normalization approaches. First, we observe once again the failure of the no-scaling approach. Second, the other three normalization approaches share a very similar behavior, and they are able to track $\bar{\varepsilon}_{max}$ with very good accuracy throughout the sequence. This is even more evident in all the cases where the data are more dominant than the physics in the loss function. The models with the highest physics-contribution still perform reasonably well, but we observe that the models with $w_{D} = 0.5$, $w_{P} = 0.5$ and $w_{D} = 0.0$ $w_{P} = 1.0$ consistently overshoot the true maximum value. This can be attributed to the need for better training, or may relate to inherent problems of physics-governed training of neural networks \cite{wang2021understanding, karniadakis2021physics}. In view of this, and given the promising signs from all the other models, in the next section we will exclude from our investigation the TCNs without scaling as well as those with $w_{D} = 0.5$, $w_{P} = 0.5$ and $w_{D} = 0.0$, $w_{P} = 1.0$.

\subsubsection{I-FENN solver: Full vs Modified Newton-Raphson}
\label{Section:SNT_IFENN_NR}

In this section we launch a detailed study on the I-FENN solver, with the primary objective to compare the performance of the full against the modified Newton-Raphson scheme. We denote that the stark difference between the two is that in the modified NR we do not rely on the TCN-computed term $\dfrac{\partial{\bar{\varepsilon}_{eq}}}{\partial{\varepsilon_{eq}}}$. Instead, only the predicted non-local strains are utilized in the finite element stiffness computations, as outlined in Section \ref{Section:IFENN_framework}. I-FENN is applied using trained TCNs from the previous section, accounting for all normalization techniques and for 4 combinations of data-vs-physics. The chosen networks are those trained with 2000 Adam epochs. For every combination we perform one full and one modified NR analysis. In total we perform 3 x 4 x 2 = 24 analyses with I-FENN. We underline that the investigated model here is M2-linear, which has a different mesh than the one used for the training of the networks. 

The analysis is considered complete if I-FENN reaches the loadfactor $lf = 0.8$, which corresponds to a load history with 160 increments. This is consistent with the load incrementation of $\Delta u_{D} = 0.005$. We use the convergence criterion defined in Eqn. \ref{Eqn_convergence_criterion_IFENN}, with $tol = 10^{-6}$. Here we emphasize that reducing the displacement residual by six orders of magnitude at every increment is a very strict convergence criterion, which ensures the accuracy of our solvers \cite{bharali2022robust}. The maximum number of iterations per increment is set to 200. If that value is exceeded then the analysis is terminated. In the present setup, the load application is monotonic and non-adaptive. Again this is a very strict choice, and the reason behind it is that we intend to have only equidistant increments in the generated sequences. This is a requirement that is inherited by the TCN structure, and it could potentially be relaxed in the future if a different network architecture is selected.

\renewcommand{\arraystretch}{1}
\begin{table}
\caption{I-FENN with Full and Modified Newton-Raphson on the M2-linear mesh. The numbers correspond to the final converged increment of the analysis. An analysis is considered successful (complete) if it reaches the $160^{th}$ increment.}
\label{Table:Full_vs_Modified_NewtonRaphson_IFENN} \centering
\begin{tabular}{c c| |C{1.6cm} C{1.6cm}| |C{1.6cm} C{1.6cm}| |C{1.6cm} C{1.6cm}}
    \hline
    \multirow{2}{*}{\bf{Data}} & \multirow{2}{*}{\bf{Physics}} & \multicolumn{2}{c||}{\bf{Constant Decimal}} & \multicolumn{2}{c||}{\bf{Varying Multiplication}} & \multicolumn{2}{c}{\bf{Min-max}} \\
    & & \bf{Full} & \bf{Modified} & \bf{Full} & \bf{Modified} & \bf{Full} & \bf{Modified} \\
    \hline
    1.0 & 0.0 & {\bf{160}} & {\bf{160}} & 116 & {\bf{160}} & 154 & {\bf{160}}  \\
    0.9 & 0.1 & 141 & {\bf{160}} & 133 & {\bf{160}} & 138 & {\bf{160}}  \\    
    0.8 & 0.2 & 134 & {\bf{160}} & 154 & {\bf{160}} & 146 & {\bf{160}}  \\
    0.7 & 0.3 & 153 & {\bf{160}} & 154 & {\bf{160}} & 134 & {\bf{160}}  \\
    \hline
\end{tabular}
\end{table} 

In Table \ref{Table:Full_vs_Modified_NewtonRaphson_IFENN} we report the last converged increment for the 24 I-FENN analyses. Our first solid observation is that I-FENN converges in all the cases where the modified Newton-Raphson scheme was employed. This holds true regardless of the data-vs-physics combination or scaling technique utilized in the TCN training. The successful completion of all these cases is one of the most important findings of this work, as it shows for the first time that I-FENN can successfully simulate the entire requested loading history. When I-FENN was invoked with the full Newton-Raphson scheme, it was able to fully complete one analysis and almost complete 4 other cases. Generally, however, its performance was distinctively substandard than its modified Newton-Raphson counterpart, which indicates that the network-generated term $\dfrac{\partial{\bar{\varepsilon}_{eq}}}{\partial{\varepsilon_{eq}}}$ was not computed with the desired accuracy. Further experimentation with the TCN training hyperparameters, such as the network dimensions or the Adam learning rate, could perhaps improve the TCN performance and, therefore, the accuracy of the queried partial derivative. This exploration, however, was not performed since, overall, these results evidently demonstrate the robustness and superiority of I-FENN with the modified Newton-Raphson scheme.

\begin{figure}[H]
	\centering
	\includegraphics[width=1\textwidth]{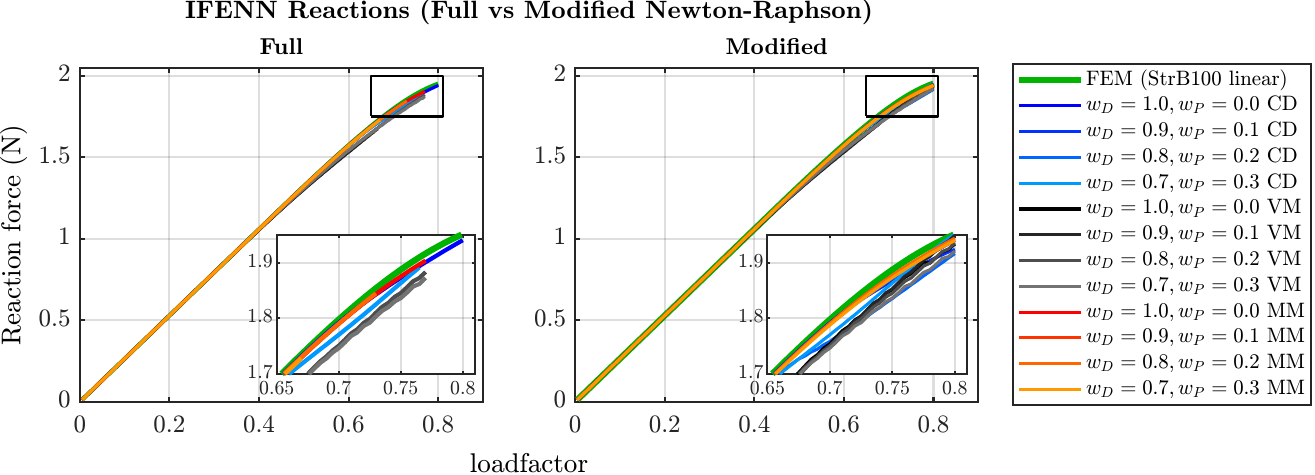}
	\caption{Reaction forces against loadfactor, using I-FENN on the M2 model with different scaling methods and data-vs-physics combinations. Comparison of the full (left graph) and the modified (right graph) Newton-Raphson schemes. Legend notation: CD - Constant Decimal, VM - Varying Multiplication, MM - Min-max normalization.}	\label{Fig:Figure_Reactions_SingleNotch_FEM_IFENN_Full_vs_Modified}
\end{figure}

In order to monitor the accuracy of I-FENN, we first plot the reaction-loadfactor curves in Fig. \ref{Fig:Figure_Reactions_SingleNotch_FEM_IFENN_Full_vs_Modified}a and \ref{Fig:Figure_Reactions_SingleNotch_FEM_IFENN_Full_vs_Modified}b, for full and modified NR respectively. Evidently, all the curves are placed very close to the true FEM curve, with a) all the MM models and b) the CD data-only model showing the best approximation. The models trained with VM show a tendency to compute slightly smaller reaction forces, which in turn implies slightly larger values in the computed non-local strains. 

\begin{figure}[H]
	\centering
	\includegraphics[width=0.9\textwidth]{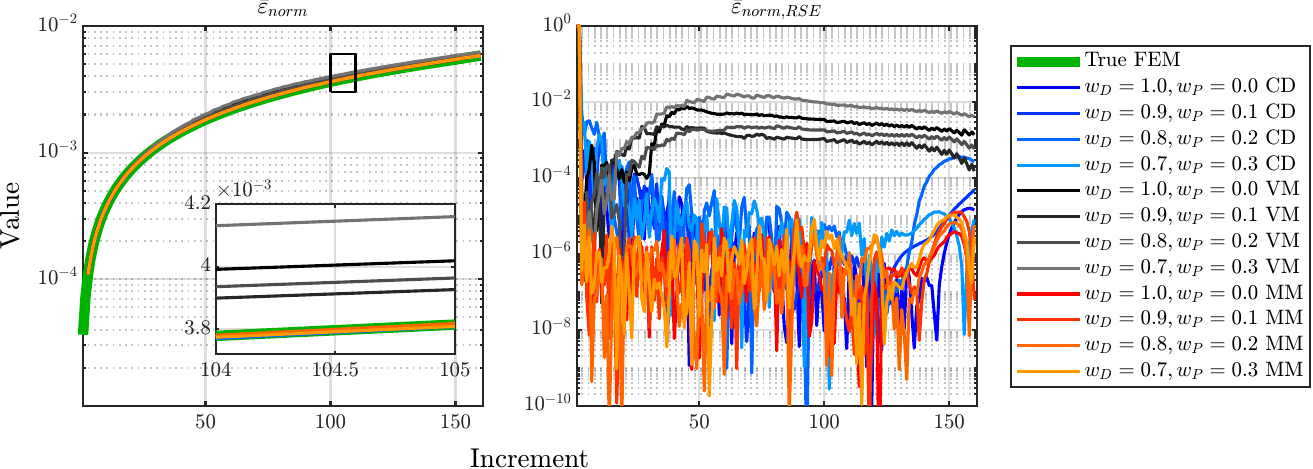}
	\caption{{\bf{a.}} Evolution of the norm of the predicted non-local strain field throughout the load history, for different scaling and data-vs-physics combinations. {\bf{b.}} Relative squared error of the norm of the predicted strains measured against the true field. Legend notation: CD - Constant Decimal, VM - Varying Multiplication, MM - Min-max normalization.}
	\label{Fig:Figure_enonlocal_norm_max_rse_SingleNotch_FEM_IFENN}
\end{figure}

To further verify these observations we compute the norm of the non-local strain field at each load increment, and in Fig. \ref{Fig:Figure_enonlocal_norm_max_rse_SingleNotch_FEM_IFENN}a we plot these curves for both the FEM (true) and I-FENN with modified NR. This figure shows the excellent agreement between the true and predicted non-local strain norms throughout the entire load history, and the zoomed-in graph further verifies that the VM curves lie slightly above the true one. The magnitude of the difference between all FEM and I-FENN is quantified in Fig. \ref{Fig:Figure_enonlocal_norm_max_rse_SingleNotch_FEM_IFENN}b, where the relative squared error of the I-FENN non-local strain norms is plotted. This graph evidently demonstrates the level of accuracy that was obtained for all models, with the min-max and decimal networks showing the best performance.

\begin{figure}
	\centering
	\includegraphics[width=0.7\textwidth]{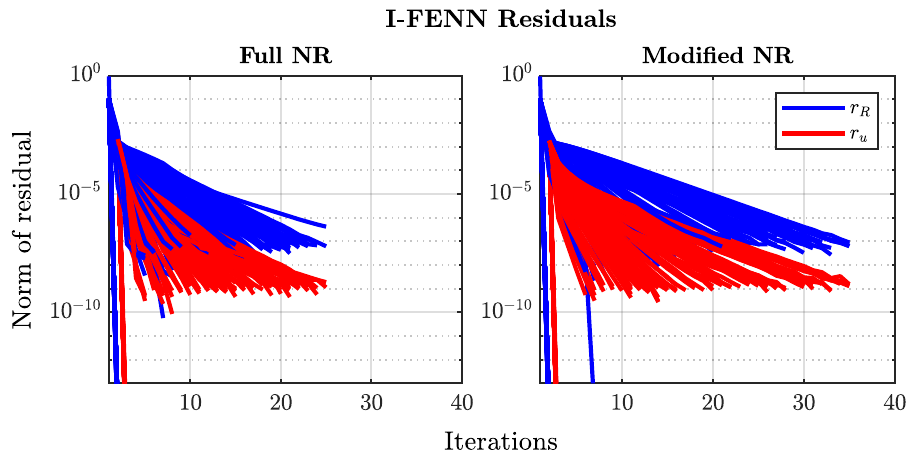}
	\caption{Evolution of the I-FENN residuals with full (left graph) and modified (right graph) Newton-Raphson on the M2 model. Blue lines indicate the residual of internal stresses, and red lines represent the residual of the displacement incremental change.}
	\label{Fig:Figure_Residuals_SingleNotch_IFENN_FNR_MNR}
\end{figure}

The final step is to examine the sanity of the I-FENN solver, and for this purpose we inspect the behavior of the residuals during convergence. Since the model trained with $w_{D} = 1.0$, $w_{P} = 0.0$ and decimal scaling was the only one that completed both the full and modified NR analysis, we select that model and for both cases we plot the residuals across all the load increments in Fig. \ref{Fig:Figure_Residuals_SingleNotch_IFENN_FNR_MNR}. The red curves correspond to the residual of the nodal displacements $r_{u}$ (Eqn. \ref{Eqn_convergence_criterion_IFENN}), and the blue curves correspond to the residual of internal stresses $r_{R}$ (Eqn. \ref{Eqn_convergence_criterion_IFENN_secondary}). We observe that all the residuals follow a steady monotonic decrease throughout both analyses, which is clear evidence of the solver health. As expected, we also note that the modified NR requires more iterations per increment than the full NR. A similar picture is observed for all the investigated models, and the reader is referred to Appendix \ref{Appendix:IFENN_Residuals} to observe the I-FENN residual minimization for all these models. 

Conclusively, the results of this section have demonstrated  that: 

\begin{itemize}

    \item I-FENN can successfully simulate the entire requested load history analysis.

    \item The modified NR configuration is a more reliable candidate than the full NR, even though the latter can also be successfully implemented. We also note that it is also more computationally efficient in the sense that the additional step of computing $\dfrac{\partial{\bar{\varepsilon}_{eq}}}{\partial{\varepsilon_{eq}}}$ with the TCN is omitted.
    
    \item Monitoring of the reaction forces and non-local strains evolution illustrates the excellent level of accuracy obtained with I-FENN.
       
    \item Both solvers exhibit a healthy reduction in the residuals across the entire simulation.
    
\end{itemize}

\subsubsection{Computational savings against benchmark FEM}
\label{Section:SNT_savings}

In this section, we explore the computational gains from I-FENN. Informed by our previous investigation, we choose the TCN with $w_D = 1.0$, $w_P = 0.0$, decimal scaling and 1000 Adam epochs. We use the modified Newton-Raphson scheme, and we apply I-FENN on the test models that are reported in Table \ref{Table:SNT_geometries}. We then perform an FEM simulation on the same models using the benchmark monolithic and staggered solvers and compare their computational performance with I-FENN. We underline that for I-FENN we account for all sources of computational expense: data generation using the M1-quad model, TCN training and FEM analysis on the testing model. The conventional FEM simulations as well as the FEM parts of I-FENN are conducted with our CPU-operated in-house MATLAB solver. The TCN training and the TCN iterative predictions of I-FENN are conducted using Python and GPU parallelization, harnessing therefore the readily available computational advantages offered by neural networks. We note that our current implementation requires a recursive file exchange between MATLAB and Python. Since this is just an artifact of our coding implementation and it is not pertinent to the overall algorithm logic, this step is not accounted in the total cost measurements. All the code and data will be made publicly available upon publication of the article to allow reproducibility of our results. The FEM simulations were performed on a Dell workstation with Intel(R) Xeon(R) W-2223 CPU @ 3.60 GHz, 64GB RAM, NVIDIA Quadro P1000 GPU, and the TCN operations (training and predictions) were performed on a Dell laptop with Intel(R) Core(TM) i7-10750H CPU @ 2.60GHz, 32.0 GB RAM, and NVIDIA GeForce GTX 1650 Ti GPU. 

The results of our analysis are presented in Table \ref{Table:Computational_Savings_SNT}. First, we mention that I-FENN can successfully complete the simulation of all the testing models. This holds true regardless of the element order or the number of Gauss points in the test model, which further confirms the validity of the overarching algorithm. Second, we observe that in all cases I-FENN is outperforming the two conventional FEM solvers, reducing the total time by a factor around $12\%-23\%$ depending on the problem. This sign provides initial yet clear evidence of the ability of our approach to accelerate the numerical solution of non-local gradient damage propagation, which is one of the primary objectives of our work. We also remark that the computational efficiency of I-FENN tends to increase as the mesh resolution is more refined, which is more apparent if one compares the computational performance between the M2-quad and M3-quad models. A more in-depth evaluation of the performance of I-FENN against the conventional solvers is conducted in the validation models presented next.

\renewcommand{\arraystretch}{1}
\begin{table}[t]
\caption{Comparison of computational performance between the three solvers for the single notch tension problem: I-FENN, FEM monolithic and FEM staggered. I-FENN has three-time sources: data generation on M1-quad, network training, and FEM analysis on the queried model. Time units are seconds.}
\label{Table:Computational_Savings_SNT} \centering
\begin{tabular}{c|| C{1.4cm} c c C{1.4cm}|| C{1.4cm} C{1.4cm}|| C{1.4cm} C{1.4cm}}
    \hline
    \multirow{2}{*}{\bf{Model}} & \multicolumn{4}{c||}{\bf{I-FENN}} & \multicolumn{2}{c||}{\bf{FEM monolithic}} & \multicolumn{2}{c}{\bf{FEM staggered}} \\
    & \bf{FEM} & \bf{M1-quad} & \bf{TCN} & \bf{Total} & \bf{Total} & \bf{Savings} & \bf{Total} & \bf{Savings} \\
    \hline
    M4-linear & 18202 & 607 & 1050 & 19859 & 22928 & {\bf{13\%}} & 22579 & {\bf{12\%}} \\
    M2-quad & 8961 & 607 & 1050 & 10618 & 12240 & {\bf{13\%}} & 12750 & {\bf{17\%}} \\ 
    M3-quad & 27252 & 607 & 1050 & 28909 & 37465 & {\bf{23\%}} & 36694 & {\bf{21\%}} \\
    \hline    
\end{tabular}
\end{table}

\subsection{Validation Model 1: Double notch under tension}
\label{Section:DNT}

\begin{figure}[H]
    \centering
    \includegraphics[width=0.6\textwidth]{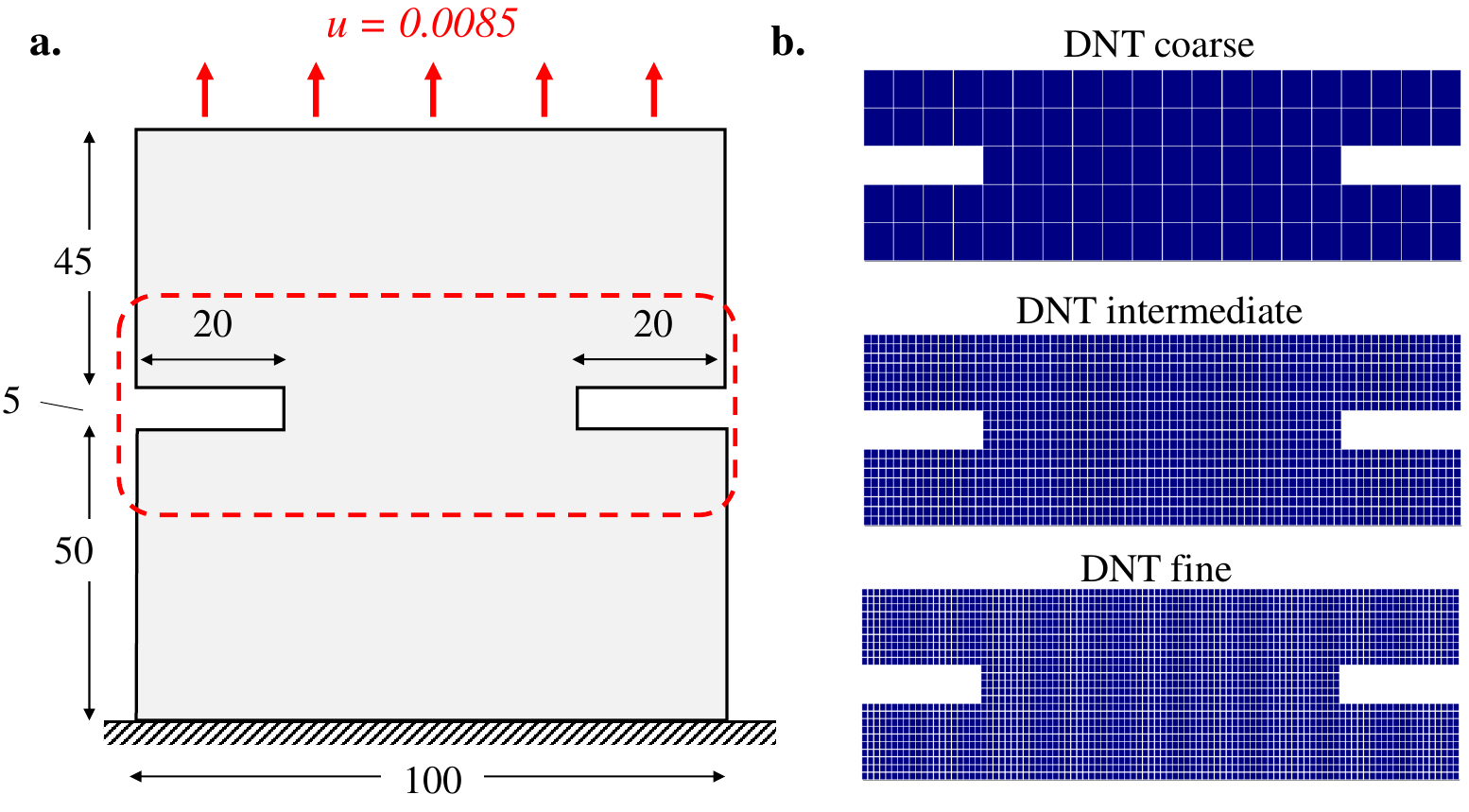}
    \caption{{\bf{a.}} Schematic graph of the double notch under tension (DNT) problem: geometry, loading, and boundary conditions. {\bf{b.}} Three mesh resolutions are generated: the DNT coarse model is used for the TCN training, and I-FENN is implemented on DNT intermediate and DNT fine.}
    \label{Fig:Figure_DNT_Geometry}
\end{figure} 

The first validation model is a 2D double-notch under tension (DNT) domain. A schematic graph of its geometry and boundary conditions is shown in Fig. \ref{Fig:Figure_DNT_Geometry}a. The bottom edge is fixed in both directions, and the top edge is pulled upwards with a prescribed displacement. The same characteristic length and Mazar's law parameters as in the SNT problem are used. Following \cite{wang2023localizing, pantidis2023integrated}, the equivalent strain is defined here as  $\varepsilon_{eq} = \frac{k-1}{2k(1-2\nu)} + \frac{1}{2k} \sqrt{\frac{(k-1)^2}{(1-2\nu)^{2}}I_{1}^{2} + \frac{2k}{(1+\nu)^2}J_{2}}$, where $I_{1} = tr(\boldsymbol{\varepsilon})$, $J_{2} = 3tr(\boldsymbol{\varepsilon} \cdot \boldsymbol{\varepsilon}) - tr^{2}(\boldsymbol{\varepsilon})$ are the strain invariants. Plane strain conditions are assumed, and three mesh idealizations are used. The $DNT \ Coarse$ model has quadratic square elements of size $l_{elem} = 5.0 \ mm$ and it is used for the training data generation. The $DNT \ Intermediate$ and $DNT \ Fine$ models have linear square elements with $l_{elem} = \{1.25, 1.0\} \ mm$ respectively, and they are used as the testing models. The Coarse, Intermediate and Fine models have 3528, 25088 and 39200 Gauss points respectively, and a close-up view of each mesh discretization is shown in Fig. \ref{Fig:Figure_DNT_Geometry}b. The TCN is trained only with data and a decimal scaling factor of $10^{4}$, with Adam and L-BFGS capped at 2000 and 5000 epochs respectively. I-FENN is activated at the $3^{rd}$ load increment, and this is simply to demonstrate the feasibility of our code to switch between the benchmark FEM solver and I-FENN during the analysis at the user-end will. I-FENN is implemented with the modified Newton-Raphson scheme and it is compared against the monolithic and staggered FEM solvers. In all cases the load is incremented until $lf = 0.8$ with a constant step of $0.005$. The FEM computations were performed on a Dell G5 laptop with Intel(R) Core(TM) i7-10750H CPU @ 2.59GHz, 16.0 GB RAM, NVIDIA GeForce RTX 2060 GPU, and the TCN operations (training and predictions) were performed on a Dell laptop with Intel(R) Core(TM) i7-10750H CPU @ 2.60GHz, 32.0 GB RAM, and NVIDIA GeForce GTX 1650 Ti GPU.
 
\renewcommand{\arraystretch}{1}
\begin{table}[H]
\caption{Comparison of computational performance between the three solvers for the double notch tension problem: I-FENN, FEM monolithic and FEM staggered. Time units are seconds.}
\label{Table:Computational_Savings_DNT} \centering
\begin{tabular}{c|| C{1.4cm} c c C{1.4cm}|| C{1.4cm} C{1.4cm}|| C{1.4cm} C{1.4cm}}
    \hline
    \multirow{2}{*}{\bf{Model}} & \multicolumn{4}{c||}{\bf{I-FENN}} & \multicolumn{2}{c||}{\bf{FEM monolithic}} & \multicolumn{2}{c}{\bf{FEM staggered}} \\
    & \bf{FEM} & \bf{Coarse} & \bf{TCN} & \bf{Total} & \bf{Total} & \bf{Savings} & \bf{Total} & \bf{Savings} \\
    \hline
    Intermediate & 20152 & 611 & 434 & 21197 & 31834 & {\bf{33\%}} & 28416 & {\bf{25\%}} \\
    Fine & 79821 & 611 & 434 & 80866 & 131582 & {\bf{39\%}} & 117047 & {\bf{31\%}} \\
    \hline    
\end{tabular}
\end{table} 

The computational performance of the three solvers is shown in Table \ref{Table:Computational_Savings_DNT}, and time is measured in the same fashion as in the SNT model. Here I-FENN shows greater computational savings than in the SNT model, being 33-39\% faster than the monolithic solver and 25-31\% faster than the staggered approach. This is a prominent reduction in the computational expense, which further demonstrates the computational advantages of I-FENN. Also, similar to the SNT case, the efficiency of our framework increases as the size of the FEM mesh is increasing. This observation leads to the conclusion that I-FENN is particularly well-suited for problems where a very high mesh resolution is required, since these cases allow for its potential to be harnessed to a greater extent. 

\begin{figure}[H]
	\centering
	\includegraphics[width=0.95\textwidth]{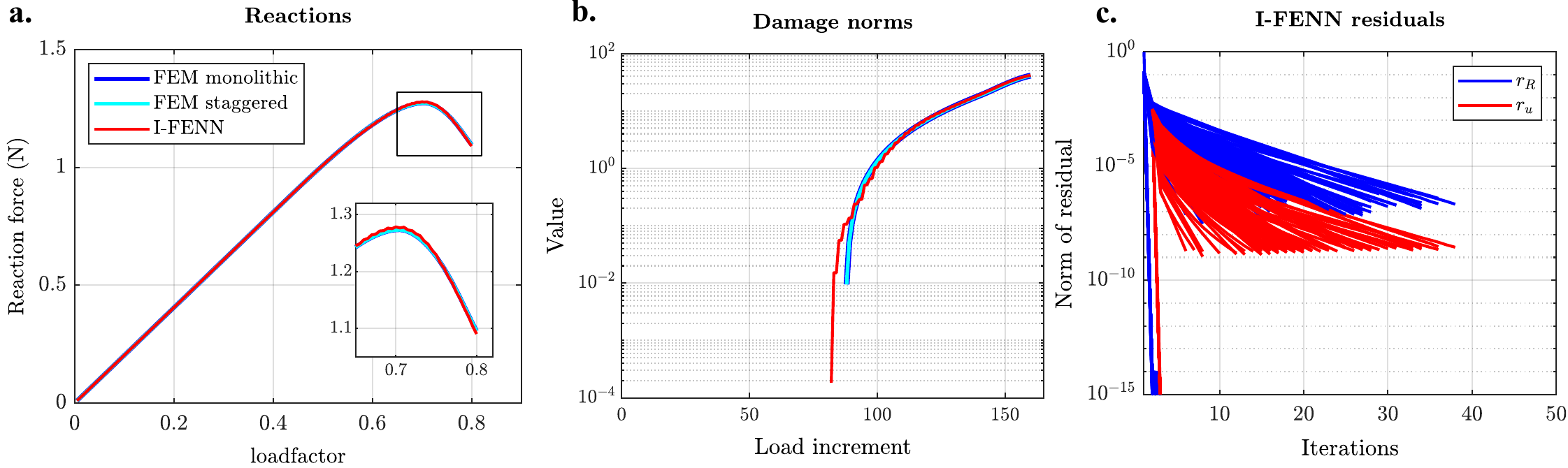}
	\caption{DNT Fine model: {\bf{a.}} Reaction-loadfactor curves and {\bf{b.}} second norm of the non-local damage profile for the three solvers. {\bf{c.}} Residual minimization in the I-FENN analysis.}
	\label{Fig:Figure_DNT_reactions_damagenorm_residuals}
\end{figure}

\begin{figure}[H]
	\centering
	\includegraphics[width=0.8\textwidth]{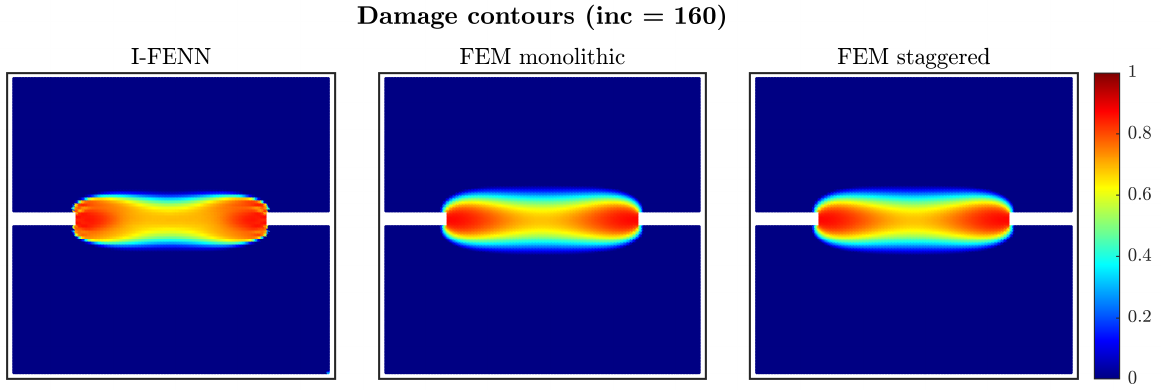}
	\caption{DNT Fine model: damage contours at the last increment of the analysis for I-FENN (left), FEM monolithic (middle), and FEM staggered (right).}
	\label{Fig:Figure_DNT_E100_Damage_Predictions_FEM_IFENN}
\end{figure}

Additionally, we examine the accuracy of I-FENN against the benchmark FEM solvers by comparing a) the reaction force-displacement curves and b) the second norm of the non-local damage fields throughout the load history for the DNT Fine model. These curves are plotted in Fig. \ref{Fig:Figure_DNT_reactions_damagenorm_residuals}a and Fig. \ref{Fig:Figure_DNT_reactions_damagenorm_residuals}b respectively, and their exceptional match further verifies the accuracy of I-FENN. Fig. \ref{Fig:Figure_DNT_E100_Damage_Predictions_FEM_IFENN} shows the non-local damage contours at the last increment of the analysis for the three solvers, where a very good resemblance between I-FENN and FEM can be observed. As a final check of our framework's performance, we present the I-FENN residuals for the DNT Fine model in Fig. \ref{Fig:Figure_DNT_reactions_damagenorm_residuals}c. These curves follow a monotonically decreasing trend throughout all the increments of the load history, which provides additional confidence in the sanity of the I-FENN solver. Similar observations for all these patterns hold true for the DNT Intermediate model.

\subsection{Validation Model 2: Single notch under shear (SNS)}
\label{Section:SNS}

\begin{figure}[H]
    \centering
    \includegraphics[width=0.87\textwidth]{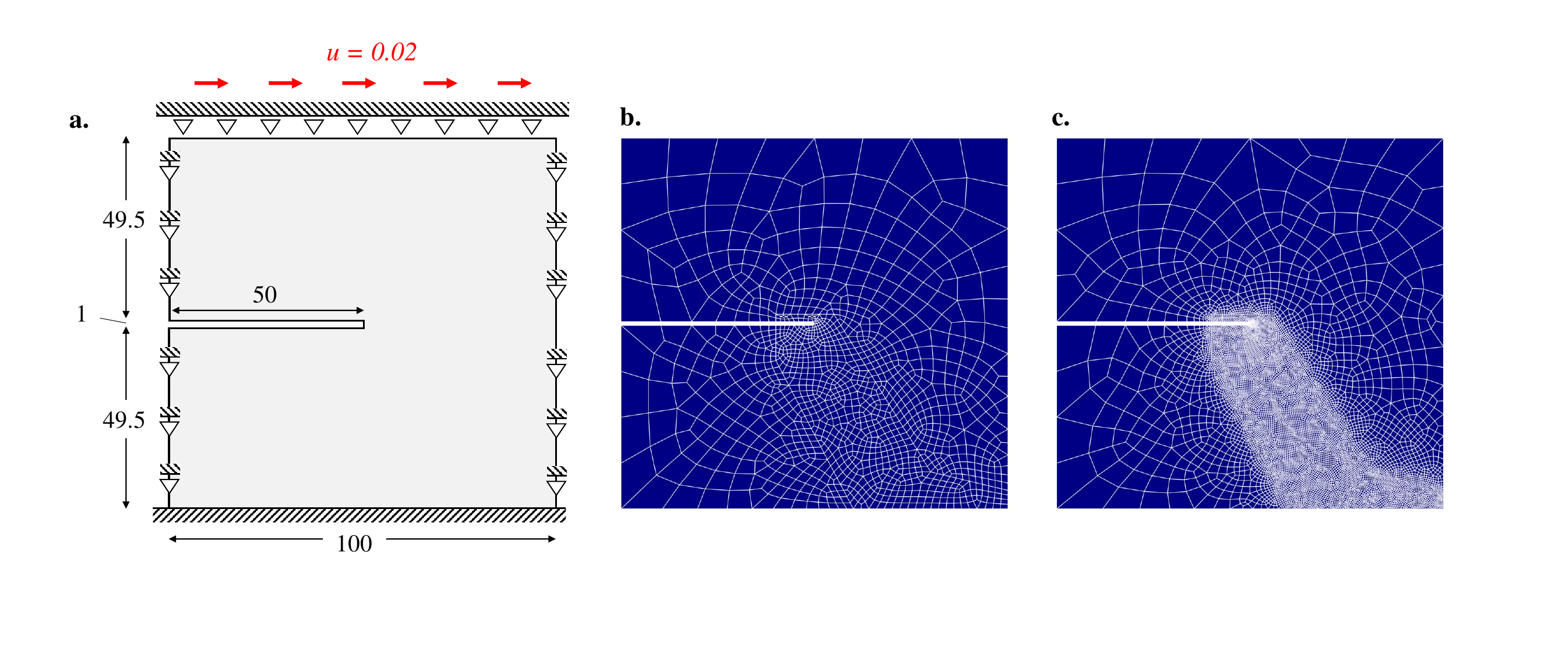}
    \caption{{\bf{a.}} Schematic graph of the single notch under shear (SNS) problem: geometry, loading, and boundary conditions. Two mesh resolutions are used for this problem: {\bf{b.}} coarse (TCN training) and {\bf{c.}} fine (I-FENN implementation).}
    \label{Fig:Figure_SNS_Geometry}
\end{figure}

The second validation model is a 2D square domain with a single notch loaded with a shear displacement (SNS). This is another benchmark example that is studied extensively in the literature \cite{deborstcomparison, kristensen2020phase}. The geometric, loading and boundary conditions of the model are shown in Fig. \ref{Fig:Figure_SNS_Geometry}a. An unstructured mesh is used for this problem, with two mesh idealizations: the $Coarse$ mesh (5104 GPs) shown in Fig. \ref{Fig:Figure_SNS_Geometry}b is used to generate the TCN training dataset, and the $Fine$ mesh (39300 GPs) in Fig. \ref{Fig:Figure_SNS_Geometry}c is the testing model. The equivalent strain definition, Mazars' damage law parameters, TCN training configuration, and I-FENN/FEM loading hyperparameters are the same as in the DNT problem. All simulations were performed on a Dell laptop with Intel(R) Core(TM) i7-10750H CPU @ 2.60GHz, 32.0 GB RAM, NVIDIA GeForce GTX 1650 Ti GPU.

\begin{figure}
    \centering
    \includegraphics[width=1\textwidth]{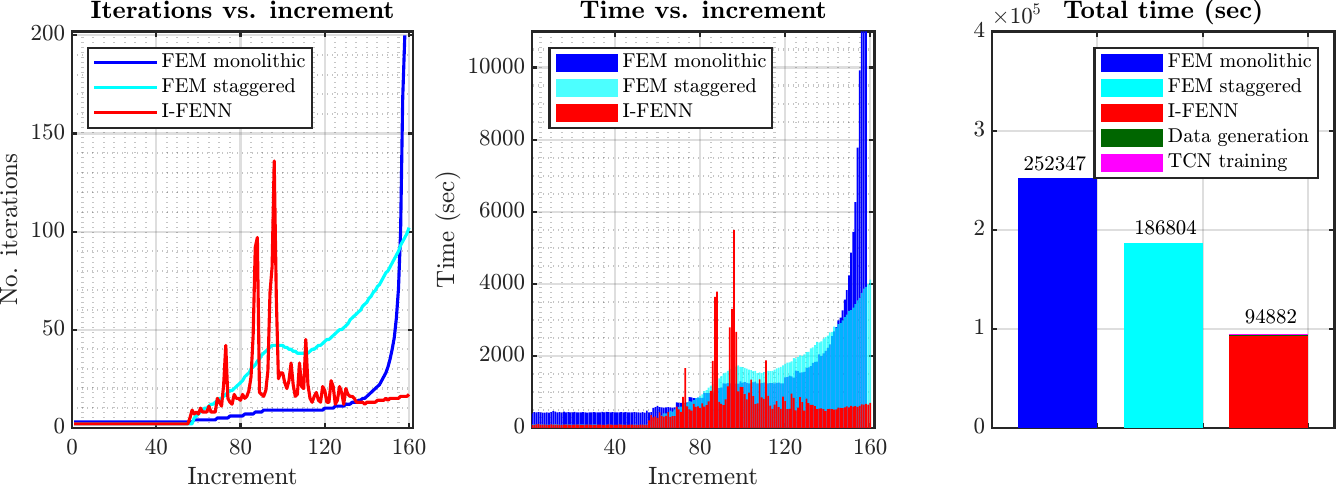}
    \caption{Comparison of the computational cost between I-FENN and FEM for the SNS model, using $tol = 10^{-6}$: {\bf{a.}} Simulation time per load increment. {\bf{b.}} Total time comparison. IFENN is almost $60\%$ faster compared to the benchmark monolithic FEM solution. In the former we account all sources of computational expense: coarse mesh FEM analysis, TCN training, and I-FENN simulation of the test model.}
    \label{Fig:Figure_SNS_Time_conv1e6}
\end{figure}

\begin{figure}[H]
    \centering
    \includegraphics[width=0.9\textwidth]{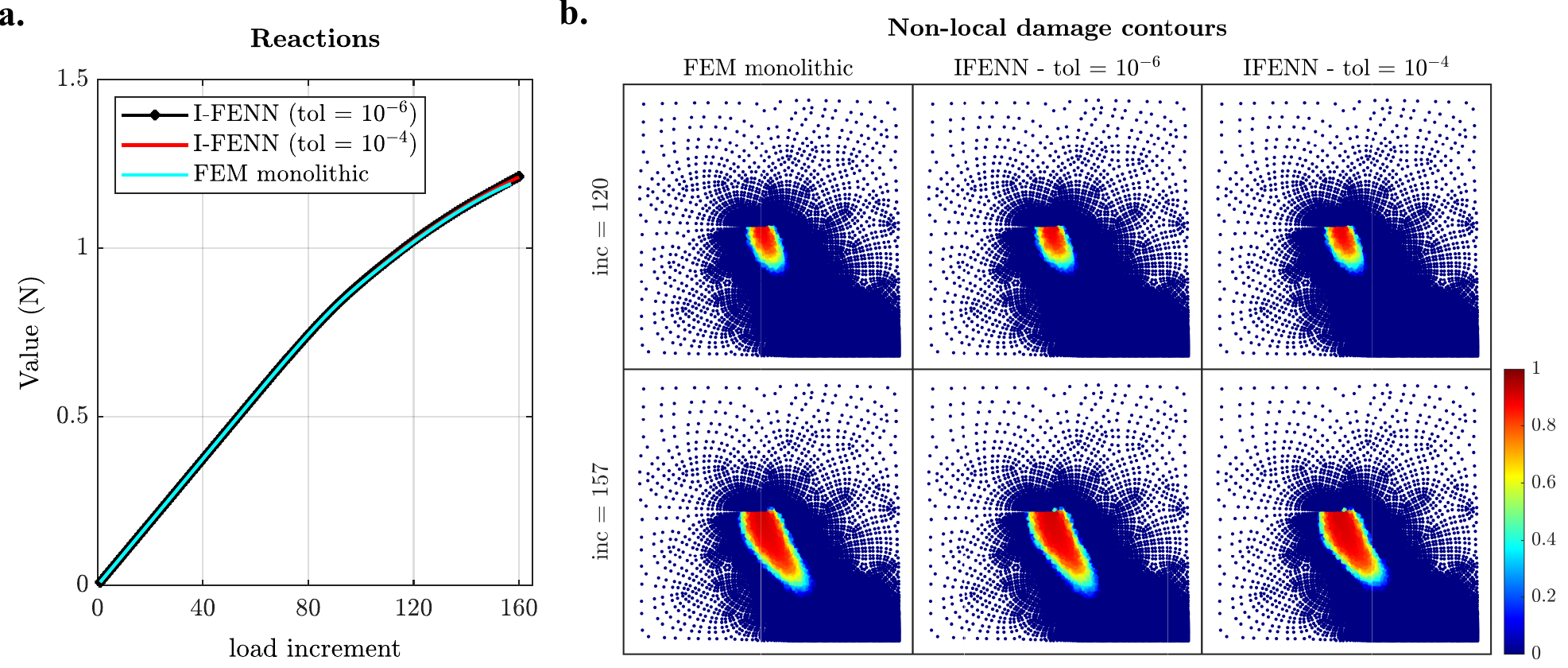}
    \caption{{\bf{a.}} Reaction forces for the SNS model for I-FENN ($tol = 10^{-6}$ and $tol = 10^{-4}$), and FEM monolithic ($tol = 10^{-6}$). The excellent agreement of the curves shows that the reduction in the convergence tolerance had a negligible impact in the accuracy of the model. {\bf{b.}} Damage contours at the $120^{th}$ and $157^{th}$ increment for the three cases.}
    \label{Fig:Figure_SNS_FEM_IFENN_reactions_damage_comparison}
\end{figure}

Fig. \ref{Fig:Figure_SNS_Time_conv1e6} displays the time comparison between the three solvers for the SNS model. The left graph shows the number of iterations per increment for I-FENN (red), FEM monolithic (blue) and FEM staggered (transparent cyan), the middle graph shows the simulation time of each increment, and the right graph shows the total time comparison.  Several interesting remarks can be drawn here:

\begin{itemize}

    \item First, we emphasize that monolithic FEM was not able to complete the analysis in the requested sequence since it exceeded the maximum number of iterations allowed for convergence on the $158^{th}$ increment. On the contrary I-FENN was able to do so, which evidently shows the strength of the proposed framework and its ability to move further across the equilibrium path. 

    \item We notice that as damage is initiated and we enter the initial stages of the non-linear regime, both I-FENN and FEM staggered undergo a drastic increase in the number of iterations/increment, and consequently their time/increment. This behavior is expected for staggered approaches due to the drastic increase in the number of iterations for fixed-step increments \cite{gerasimov2016line}. Since these two numerical methods are conceptually similar, this trend is also a sanity sign for I-FENN.
    
    \item The presence of spikes in a few I-FENN increments is expected and it can be explained as follows. This is an byproduct of two competing factors: the need to satisfy such a low tolerance ($tol = 10^{-6}$), while using an imperfect TCN. The networks that are utilized within I-FENN are the product of an optimization process that has most probably arrived to a local minimum \cite{pantidis2023116160, mishra2022estimates}. Consequently, a trained TCN is not guaranteed to capture perfectly the true field in every increment throughout the entire load history, and therefore it is reasonable to expect in some increments more iterations until convergence is achieved. This is particularly the case when such a low tolerance has to be satisfied, and the impact of $tol$ will be discussed later in greater detail.
    
    \item The time/increment for both FEM solvers follows an increasing trend, and particularly for the monolithic it grows exponentially towards the end of the analysis. On the contrary, the I-FENN time/increment stays rather constant after the $120^{th}$ increment, which shows that I-FENN is able to converge faster in the non-linear region of the load sequence. This is attributed to the already acquired knowledge of the non-local strain profile in that zone from the Coarse mesh analysis, which is feeding the I-FENN analysis of the Fine mesh through the trained TCN. This is the primary source of the substantial computational efficiency of I-FENN. 
    
    \item As seen in Fig. \ref{Fig:Figure_SNS_FEM_IFENN_reactions_damage_comparison}b, I-FENN is $\approx$ $62\%$ and $49\%$ faster than the FEM monolithic and staggered schemes respectively for the SNS problem investigated here. This marks the biggest time reduction reported so far. It is also very interesting to observe how small is the contribution of the Coarse FEM analysis and the TCN training in the total I-FENN time, with both sources cumulatively constituting $\approx \ 2.5\%$ of the total I-FENN cost. 
    
\end{itemize}

\begin{figure}[H]
    \centering
    \includegraphics[width=1\textwidth]{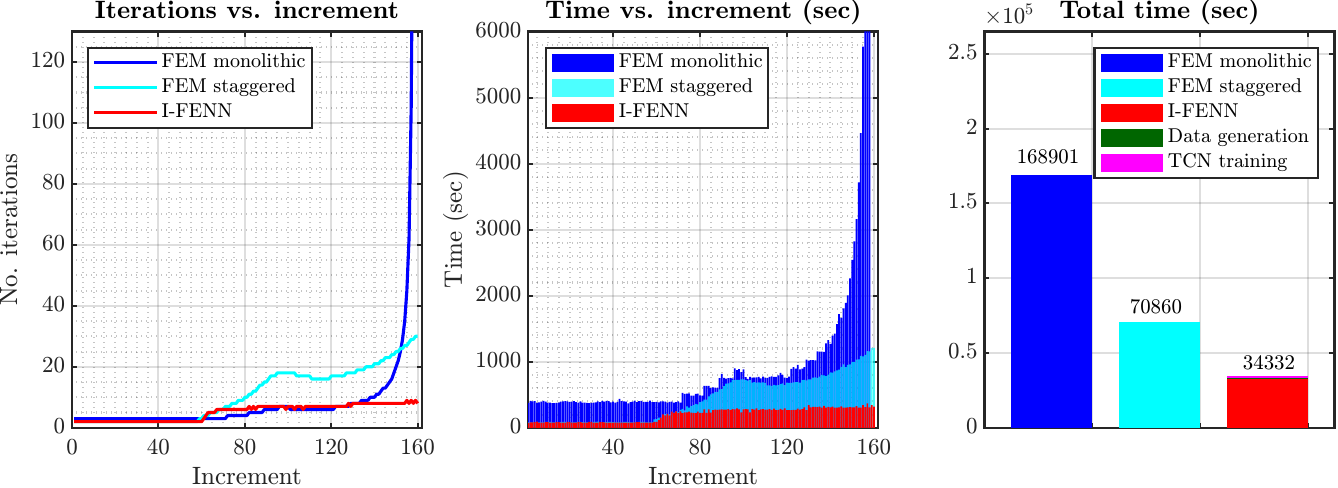}
    \caption{Comparison of computational performance for the SNS model using $tol = 10^{-4}$, following an identical layout as in Fig. \ref{Fig:Figure_SNS_Time_conv1e6}}
    \label{Fig:Figure_SNS_Time_conv1e4}
\end{figure}

So far, our goal was to demonstrate the robustness of our solver, and thus, we have imposed a very strict convergence criterion by setting $tol = 10^{-6}$. To further elucidate the impact of this constraint, we now slightly relax its value and we repeat all the SNS simulations using $tol = 10^{-4}$. While this is still considered a hard requirement and ensures numerical accuracy, this variation allows us to explore deeper its impact on the performance of all solvers. In Fig. \ref{Fig:Figure_SNS_Time_conv1e4} we present the results of this investigation, and this figure follows an identical layout to Fig. \ref{Fig:Figure_SNS_Time_conv1e6}.

The first evident observation is the absence of spikes in the iterations vs. increment graph. This corroborates our previous argument and provides further evidence of the impact of the very low tolerance on the sudden increase of iterations in some load increments. Also, the time vs. increment trends remain the same for the FEM solvers, and we once again confirm the ability of I-FENN to maintain the same number of iterations vs. increments throughout the inelastic regime. All these factors ultimately contribute to the highest reduction in the computational time that we report in this work, which is around $80\%$ and $52\%$ against the benchmark FEM monolithic and staggered schemes, respectively. 

To ensure that the accuracy of our numerical solution is not compromised by the slightly relaxed convergence criterion, we plot in Fig. \ref{Fig:Figure_SNS_FEM_IFENN_reactions_damage_comparison} the reaction forces of the model using: a) FEM monolithic with $tol = 10^{-6}$, which we consider as "true", b) I-FENN with $tol = 10^{-6}$ and c) I-FENN with $tol = 10^{-4}$. Excellent coincidence between the three curves can be observed throughout the entire load history. The damage contours of the $120^{th}$ and $157^{th}$ increment are also shown in Fig. \ref{Fig:Figure_SNS_FEM_IFENN_reactions_damage_comparison}b, and these graphs provide additional evidence of the comparative accuracy between these approaches. 

\subsection{Extension of I-FENN using hyperparameter optimization}
\label{Sec:I-FENN_Hyperopt}

\begin{figure}[H]
    \centering
    \includegraphics[width=0.95\textwidth]{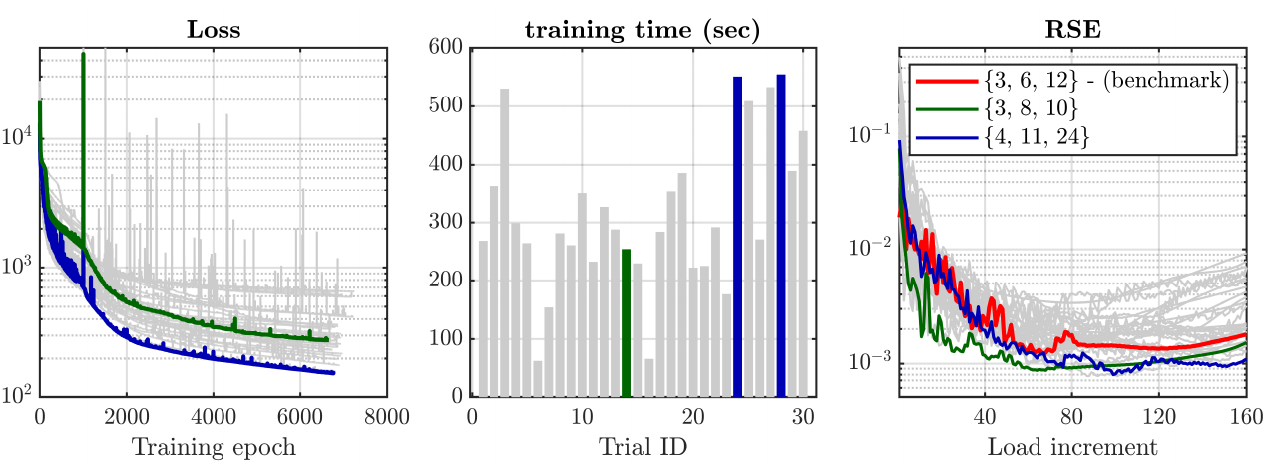}
    \caption{Hyper-parameter optimization: relative squared error of the predictions using the benchmark and optimized TCNs.}
    \label{Fig:Figure_Hyperopt}
\end{figure}

In this section, we illustrate how a hyper-parameter optimization search can potentially boost the performance of I-FENN. Using the SNS model, we implement the $Hyperopt$ package \cite{bergstra2013making} and we explore the three size-related hyperparameters of the network: number of dilations $dil$, number of filters $num_{filters}$ and kernel size $k_{size}$. The ranges of the three hyperparameters are [2-4], [2-12], and [2-24], respectively, resulting in a total number of 759 potential different sizes. We perform 30 trials with decimal scaling, 1000 Adam epochs and 5000 L-BFGS epochs. Due to the large memory requirements, these networks were trained on a High-Performance Computing cluster. Fig. \ref{Fig:Figure_Hyperopt} presents the results of this investigation: Fig. \ref{Fig:Figure_Hyperopt}a shows the evolution of the training loss ${\boldsymbol{\mathcal{L}}}$ for all 30 models, Fig. \ref{Fig:Figure_Hyperopt}b shows their training times, and Fig. \ref{Fig:Figure_Hyperopt}c presents the relative squared error between the predictions and the true non-local strain field, $\bar{\varepsilon}_{RSE}$. For the sake of comparison, we also plot ${\boldsymbol{\mathcal{L}}}$ and $\bar{\varepsilon}_{RSE}$ of our benchmark TCN with red color, and we recall that this network was trained with 2000 Adam epochs. Based on inspection of Fig. \ref{Fig:Figure_Hyperopt} we can make the following remarks.

The "best" network, i.e., the one with the consistently smaller loss values throughout the entire training history, is denoted with blue in Fig. \ref{Fig:Figure_Hyperopt} and it has a size of $dil = 4$, $num_{filters} = 11$ and $k_{size} = 24$. This size essentially corresponds to the upper bounds of the search space. This finding is in accordance with the universal approximation theorem, which states that as the network increases in size its approximation capability increases as well \cite{pantidis2023116160, shin2020convergence, mishra2022estimates, chen1995universal}. It is also very interesting that out of the 759 combinations, Hyperopt arrived at that size twice within 30 attempts, at trials 24/30 and 28/30. As seen in Fig. \ref{Fig:Figure_Hyperopt}c, this network yields one of the lowest $\bar{\varepsilon}_{RSE}$ curves below the benchmark model. At the same time, we notice the presence of the network at trial 14/30, which has a size of $dil = 3$, $num_{filters} = 8$, and $k_{size} = 10$. The size of this network is remarkably close to our benchmark model, but that TCN a) yields a distinctively lower $\bar{\varepsilon}_{RSE}$ than the benchmark, and b) requires half the training time of the "best" one. In fact, the "best" model was not even trainable on the local machine that we trained the benchmark TCN, since its size requirements exceeded the available memory. Therefore, it is evidently shown that a hyper-parameter optimization search can identify a network with better performance than what we have investigated so far, potentially improving even more the reported computational savings, while also avoiding the computationally expensive or even infeasible large network sizes. It is anticipated that a more detailed investigation will lead to an even more substantial improvement, but such an extension is outside of our main scope.

\section{Summary and Conclusions}
\label{Section:Discussion_Conclusions}

In this paper, we propose a TCN-based formulation of I-FENN for the load history analysis of non-local gradient damage propagation. We delve with detail into several critical components of this setup: the I-FENN workflow with TCNs, the choice of the I-FENN solver, the TCN training configuration, and the necessity of normalization and un-normalization schemes. The results of our work demonstrated for the first time the ability of our framework to simulate accurately the history of damage propagation across several benchmark examples, and demonstrated its computational savings against classical FEM approaches. Very strict convergence criteria were satisfied at every increment of the analysis in order to ensure the robustness of our numerical approach. From a computational efficiency standpoint, I-FENN showed a remarkable speedup compared to the conventional FEM solvers (both monolithic and staggered schemes) for all problems under consideration. The computational savings showed a clear tendency to increase more as: a) the analysis proceeds further in the inelastic zone, b) larger models with finer mesh resolutions are utilized, and c) the convergence tolerance is slightly relaxed without compromise in the accuracy.


\section*{CRediT authorship statement}
\label{Section:Credit_Statement}

\textbf{Panos Pantidis:} Conceptualization, Methodology, Software, Formal Analysis, Writing - Original draft, Data Curation, Visualization, Supervision. \textbf{Habiba Eldababy:} Software, Formal Analysis, Writing - Review and Editing. \textbf{Diab Abueidda:} Software, Writing - Review and Editing. \textbf{Mostafa Mobasher:} Conceptualization, Methodology, Writing - Review and Editing, Supervision, Project administration, Funding acquisition.

\section*{Acknowledgements}
\label{Section:Acknowledgements}

This work was partially supported by the Sand Hazards and Opportunities for Resilience, Energy, and Sustainability (SHORES) Center, funded by Tamkeen under the NYUAD Research Institute. The authors would also like to acknowledge the support of the NYUAD Center for Research Computing for providing resources, services, and staff expertise.

\section*{Data availability}
\label{Section:Data_Availability}

All the code and data used in this work will be made publicly available upon publication of the article.

\newpage
\bibliography{bibliography}

\newpage
\appendix
\section{Implementation algorithms of monolithic and staggered FEM solvers}
\label{Appendix:FEM_Algorithms}

We use the following notation in the algorithms below: right subscript $n$ is the load increment, left subscript $i$ is the iteration number, $i_{max}$ is the maximum allowable number of iterations per increment, $\bf{x}$ is the vector of DOFs ($\bf{x} = [\bf{u} \; \;  \boldsymbol{\bar{\varepsilon}}]$), $u_{p}$ is the total external displacement, $lf$ is the loadfactor, right superscripts $f$ and $e$ denote the free and essential boundary. The MATLAB $mldivide$ operation \cite{MATLAB_mldivide} is used to solve all systems of equations.

\begin{algorithm}[H]
\caption{Implementation algorithm of monolithic FEM solver}
\label{Alg:Algorithm_FEM_monolithic}
\linespread{1.2}\selectfont
Initialize input; $n = 0, i = 0$ \\
\While{$lf \leq lf_{max}$}
{$n = n + 1$, $i = 1$ \\
Compute Jacobian $\bf{J}$ (Eqns. \ref{Eqn_Juu} - \ref{Eqn_Jee}), partition to free and essential parts \\
Update essential boundary: ${\bf{x}}^{e}_{n}$ = ${\bf{x}}^{e}_{n-1}$ + $\delta lf$ $\cdot$ $u_{p}$  \\ 
Compute residual $\bf{R}$ (Eqns. \ref{FEM_Weak_Res_u}, \ref{FEM_Weak_Res_e}), partition to free and essential parts \\ 
\While{true}
{Solve the system: ${\bf{J}}^{f}$ $\delta {\bf{x}}^{f}$ = - ${\bf{R}}^{f}$  (Eqn. \ref{FEM_SystemEqn_Full}) \\
${\bf{x}}^{f}$  = ${\bf{x}}^{f}$  + $\delta {\bf{x}}^{f}$  \\
Compute $\bf{R}$, $\bf{J}$ (Eqns. \ref{FEM_Weak_Res_u}, \ref{Eqn_Juu} - \ref{Eqn_Jee}), partition to free and essential parts \\
\lIf{$\dfrac{\prescript{}{i}\lVert {\bf{ \delta x}}^{f} \rVert_{2}}{\prescript{}{1}\lVert {\bf{\delta x}}^{f}\rVert_{2}} < 10^{-6}$}
{compute reactions, {\bf{break}}}
\lElse{$i = i + 1$}}}
\end{algorithm}

\begin{algorithm}[H]
\caption{Implementation algorithm of staggered FEM solver}
\label{Alg:Algorithm_FEM_staggered}
\linespread{1.2}\selectfont
Initialize input; $n = 0, i = 0$ \\
\While{$lf \leq lf_{max}$}
{$n = n + 1$, $i = 1$ \\
Compute ${\bf{J}}^{uu}$ (Eqn. \ref{Eqn_Juu}), partition to free and essential parts \\
Update essential boundary: ${\bf{u}}^{e}_{n}$ = ${\bf{u}}^{e}_{n-1}$ + $\delta lf$ $\cdot$ $u_{p}$  \\ 
Compute residual ${\bf{R}}^{u}$ (Eqn. \ref{FEM_Weak_Res_u}), partition to free and essential parts \\ 
\While{true}
{Solve the system: ${\bf{J}}^{uu,f}$ $\delta {\bf{u}}^{f}$ = - ${\bf{R}}^{u,f}$ \\
${\bf{u}}^{f}$  = ${\bf{u}}^{f}$  + $\delta {\bf{u}}^{f}$  \\
Compute ${\bf{R}}^{\bar{\varepsilon}}$, ${\bf{J}}^{\bar{\varepsilon}\bar{\varepsilon}}$ (Eqns. \ref{FEM_Weak_Res_e}, \ref{Eqn_Jee})\\
Solve the system: ${\bf{J}}^{\bar{\varepsilon}\bar{\varepsilon}}$ $\delta {\boldsymbol{\bar{\varepsilon}}}$ = - ${\bf{R}}^{\bar{\varepsilon}}$ \\
${\boldsymbol{\bar{\varepsilon}}}$ = ${\boldsymbol{\bar{\varepsilon}}}$ + $\delta {\boldsymbol{\bar{\varepsilon}}}$  \\
Compute ${\bf{R}}^{u}, {\bf{J}}^{uu}$ (Eqn. \ref{FEM_Weak_Res_u}, \ref{Eqn_Juu}), partition to free and essential parts \\
Compute norm of global residual $\prescript{}{i}\lVert {\bf{r}} \rVert_{2} = \lVert {\delta {\bf{u}}^{f}} \; ; \;  {\delta {\boldsymbol{\bar{\varepsilon}}}} \rVert_{2}$ \\
\lIf{$\dfrac{\prescript{}{i}\lVert {\bf{r}} \rVert_{2}} {\prescript{}{1}\lVert {\bf{r}} \rVert_{2}} < 10^{-6}$}
{compute reactions, {\bf{break}}}
\lElse{$i = i + 1$}}}
\end{algorithm}

\newpage
\section{Computation of spatial derivatives using shape functions}
\label{Appendix:SF_derivation}

In this appendix, we present the derivation of Eqn. \ref{Eqn:SF_derivation} and we provide the mathematical formulas for all terms involved. The derivation of the shape function derivatives has benefited from the discussion in \cite{SF_derivatives}. Let us begin by deriving the second-order derivative of $\bar{\varepsilon}_{eq}$ w.r.t the natural coordinate $\xi$:

\begin{equation}
    \dfrac{\partial^{2} \bar{\varepsilon}_{eq}}{\partial \xi^{2}} = 
    \dfrac{\partial}{\partial \xi} \left( \dfrac{\partial \bar{\varepsilon}_{eq}}{\partial \xi} \right) = 
    \dfrac{\partial}{\partial \xi} \left( \dfrac{\partial \bar{\varepsilon}_{eq}}{\partial x} \dfrac{\partial x}{\partial \xi} + \dfrac{\partial \bar{\varepsilon}_{eq}}{\partial y} \dfrac{\partial y}{\partial \xi} \right) =
    \left[ \dfrac{\partial}{\partial \xi} \left( \dfrac{\partial \bar{\varepsilon}_{eq}}{\partial x} \right) \dfrac{\partial x}{\partial \xi} + \dfrac{\partial \bar{\varepsilon}_{eq}}{\partial x} \dfrac{\partial^{2} x}{\partial \xi^{2}} \right] + 
    \left[ \dfrac{\partial}{\partial \xi} \left( \dfrac{\partial \bar{\varepsilon}_{eq}}{\partial y} \right) \dfrac{\partial y}{\partial \xi} + \dfrac{\partial \bar{\varepsilon}_{eq}}{\partial y} \dfrac{\partial^{2} y}{\partial \xi^{2}} \right]
\label{Eqn:Appendix_1}    
\end{equation}

We can write the terms $\dfrac{\partial}{\partial \xi} \left( \dfrac{\partial \bar{\varepsilon}_{eq}}{\partial x} \right)$ and $\dfrac{\partial}{\partial \xi} \left( \dfrac{\partial \bar{\varepsilon}_{eq}}{\partial y} \right)$ as:

\begin{equation}
    \dfrac{\partial}{\partial \xi} \left( \dfrac{\partial \bar{\varepsilon}_{eq}}{\partial x} \right) = 
    \dfrac{\partial}{\partial x} \left( \dfrac{\partial \bar{\varepsilon}_{eq}}{\partial x} \right) \dfrac{\partial x}{\partial \xi} + 
    \dfrac{\partial}{\partial y} \left( \dfrac{\partial \bar{\varepsilon}_{eq}}{\partial x} \right) \dfrac{\partial y}{\partial \xi} = 
    \dfrac{\partial^{2} \bar{\varepsilon}_{eq}}{\partial x^{2}} \dfrac{\partial x}{\partial \xi} + 
    \dfrac{\partial^{2} \bar{\varepsilon}_{eq}}{\partial x \partial y} \dfrac{\partial y}{\partial \xi}
\label{Eqn:Appendix_2}
\end{equation}

\begin{equation}
    \dfrac{\partial}{\partial \xi} \left( \dfrac{\partial \bar{\varepsilon}_{eq}}{\partial y} \right) = 
    \dfrac{\partial}{\partial x} \left( \dfrac{\partial \bar{\varepsilon}_{eq}}{\partial y} \right) \dfrac{\partial x}{\partial \xi} + 
    \dfrac{\partial}{\partial y} \left( \dfrac{\partial \bar{\varepsilon}_{eq}}{\partial y} \right) \dfrac{\partial y}{\partial \xi} = 
    \dfrac{\partial^{2} \bar{\varepsilon}_{eq}}{\partial x \partial y} \dfrac{\partial x}{\partial \xi} + 
    \dfrac{\partial^{2} \bar{\varepsilon}_{eq}}{\partial y^{2}} \dfrac{\partial y}{\partial \xi}
\label{Eqn:Appendix_3}
\end{equation}

Re-arranging Eqn. \ref{Eqn:Appendix_1} with the aid of Eqns. \ref{Eqn:Appendix_2} and \ref{Eqn:Appendix_3} yields:

\begin{equation}
    \dfrac{\partial^{2} \bar{\varepsilon}_{eq}}{\partial \xi^{2}} = 
    \left[ \dfrac{\partial^{2} \bar{\varepsilon}_{eq}}{\partial x^{2}} \left( \dfrac{\partial x}{\partial \xi} \right)^{2} + 
    \dfrac{\partial^{2} \bar{\varepsilon}_{eq}}{\partial x \partial y} \dfrac{\partial x}{\partial \xi} \dfrac{\partial y}{\partial \xi} + 
    \dfrac{\partial \bar{\varepsilon}_{eq}}{\partial x} \dfrac{\partial^{2} x}{\partial \xi^{2}} \right] + 
    \left[ \dfrac{\partial^{2} \bar{\varepsilon}_{eq}}{\partial x \partial y} \dfrac{\partial x}{\partial \xi} \dfrac{\partial y}{\partial \xi} + 
    \dfrac{\partial^{2} \bar{\varepsilon}_{eq}}{\partial y^{2}} \left( \dfrac{\partial y}{\partial \xi} \right)^{2} +  
    \dfrac{\partial \bar{\varepsilon}_{eq}}{\partial y} \dfrac{\partial^{2} y}{\partial \xi^{2}} \right] 
\label{Eqn:Appendix_4}    
\end{equation}

We now re-arrange Eqn. \ref{Eqn:Appendix_4} to isolate the $\left( \dfrac{\partial^{2} \bar{\varepsilon}_{eq}}{\partial x^{2}} \right)$ and $\left( \dfrac{\partial^{2} \bar{\varepsilon}_{eq}}{\partial y^{2}} \right)$ terms:

\begin{equation}
    \dfrac{\partial^{2} \bar{\varepsilon}_{eq}}{\partial \xi^{2}} - 
    \dfrac{\partial \bar{\varepsilon}_{eq}}{\partial x} \dfrac{\partial^{2} x}{\partial \xi^{2}} - 
    \dfrac{\partial \bar{\varepsilon}_{eq}}{\partial y} \dfrac{\partial^{2} y}{\partial \xi^{2}} = 
    \dfrac{\partial^{2} \bar{\varepsilon}_{eq}}{\partial x^{2}} \left( \dfrac{\partial x}{\partial \xi} \right)^{2} + 
    \dfrac{\partial^{2} \bar{\varepsilon}_{eq}}{\partial y^{2}} \left( \dfrac{\partial y}{\partial \xi} \right)^{2} + 
    2 \dfrac{\partial^{2} \bar{\varepsilon}_{eq}}{\partial x \partial y} \dfrac{\partial x}{\partial \xi} \dfrac{\partial y}{\partial \xi}
\label{Eqn:Appendix_5}    
\end{equation}

If we follow a similar approach for $\dfrac{\partial^{2} \bar{\varepsilon}_{eq}}{\partial \eta^{2}}$ and $\dfrac{\partial^{2} \bar{\varepsilon}_{eq}}{\partial \xi \partial \eta}$, we arrive at the following expressions:

\begin{equation}
    \dfrac{\partial^{2} \bar{\varepsilon}_{eq}}{\partial \eta^{2}} - 
    \dfrac{\partial \bar{\varepsilon}_{eq}}{\partial x} \dfrac{\partial^{2} x}{\partial \eta^{2}} - 
    \dfrac{\partial \bar{\varepsilon}_{eq}}{\partial y} \dfrac{\partial^{2} y}{\partial \eta^{2}} = 
    \dfrac{\partial^{2} \bar{\varepsilon}_{eq}}{\partial x^{2}} \left( \dfrac{\partial x}{\partial \eta} \right)^{2} + 
    \dfrac{\partial^{2} \bar{\varepsilon}_{eq}}{\partial y^{2}} \left( \dfrac{\partial y}{\partial \eta} \right)^{2} + 
    2 \dfrac{\partial^{2} \bar{\varepsilon}_{eq}}{\partial x \partial y} \dfrac{\partial x}{\partial \eta} \dfrac{\partial y}{\partial \eta}
\label{Eqn:Appendix_6}    
\end{equation}

\begin{equation}
    \dfrac{\partial^{2} \bar{\varepsilon}_{eq}}{\partial \xi \partial \eta} -
    \dfrac{\partial \bar{\varepsilon}_{eq}}{\partial x} \dfrac{\partial^{2} x}{\partial \xi \partial \eta} -
    \dfrac{\partial \bar{\varepsilon}_{eq}}{\partial y} \dfrac{\partial^{2} y}{\partial \xi \partial \eta} = 
    \dfrac{\partial^{2} \bar{\varepsilon}_{eq}}{\partial x^{2}} \dfrac{\partial x}{\partial \xi} \dfrac{\partial x}{\partial \eta} + 
    \dfrac{\partial^{2} \bar{\varepsilon}_{eq}}{\partial y^{2}} \dfrac{\partial y}{\partial \xi} \dfrac{\partial y}{\partial \eta} +     
    \dfrac{\partial^{2} \bar{\varepsilon}_{eq}}{\partial x \partial y} \left( \dfrac{\partial x}{\partial \xi} \dfrac{\partial y}{\partial \eta} + \dfrac{\partial x}{\partial \eta} \dfrac{\partial y}{\partial \xi} \right)
\label{Eqn:Appendix_7}
\end{equation}

Re-arranging Eqns. \ref{Eqn:Appendix_5}, \ref{Eqn:Appendix_6} and \ref{Eqn:Appendix_7} into a compact matrix format we arrive at the following system:

\begin{equation}
\begin{bmatrix} 
\left( \dfrac{\partial x}{\partial \xi} \right)^{2}                         &
\left( \dfrac{\partial y}{\partial \xi} \right)^{2}                         & 
2 \dfrac{\partial x}{\partial \xi} \dfrac{\partial y}{\partial \xi}      \\ & \\
\left( \dfrac{\partial x}{\partial \eta} \right)^{2}                        & 
\left( \dfrac{\partial y}{\partial \eta} \right)^{2}                        &
2 \dfrac{\partial x}{\partial \eta} \dfrac{\partial y}{\partial \eta}    \\ & \\
\dfrac{\partial x}{\partial \xi} \dfrac{\partial x}{\partial \eta}          & 
\dfrac{\partial y}{\partial \xi} \dfrac{\partial y}{\partial \eta}          & 
\dfrac{\partial x}{\partial \xi} \dfrac{\partial y}{\partial \eta} + \dfrac{\partial x}{\partial \eta} \dfrac{\partial y}{\partial \xi} 
\end{bmatrix} \cdot 
\begin{bmatrix}
\dfrac{\partial^{2} \bar{\varepsilon}_{eq}}{\partial x^{2}} \\ \\
\dfrac{\partial^{2} \bar{\varepsilon}_{eq}}{\partial y^{2}} \\ \\
\dfrac{\partial^{2} \bar{\varepsilon}_{eq}}{\partial x \partial y} 
\end{bmatrix} 
= 
\begin{bmatrix} 
\dfrac{\partial^{2} \bar{\varepsilon}_{eq}}{\partial \xi^{2}} - 
\dfrac{\partial \bar{\varepsilon}_{eq}}{\partial x}\dfrac{\partial^{2} x}{\partial \xi^{2}} - 
\dfrac{\partial \bar{\varepsilon}_{eq}}{\partial y}\dfrac{\partial^{2} y}{\partial \xi^{2}} \\ \\ 
\dfrac{\partial^{2} \bar{\varepsilon}_{eq}}{\partial \eta^{2}} - 
\dfrac{\partial \bar{\varepsilon}_{eq}}{\partial x}\dfrac{\partial^{2} x}{\partial \eta^{2}} - 
\dfrac{\partial \bar{\varepsilon}_{eq}}{\partial y}\dfrac{\partial^{2} y}{\partial \eta^{2}} \\ \\
\dfrac{\partial^{2} \bar{\varepsilon}_{eq}}{\partial \xi\eta} - 
\dfrac{\partial \bar{\varepsilon}_{eq}}{\partial x}\dfrac{\partial^{2} x}{\partial \xi \partial \eta} - 
\dfrac{\partial \bar{\varepsilon}_{eq}}{\partial y}\dfrac{\partial^{2} y}{\partial \xi \partial \eta}
\end{bmatrix}
\label{Eqn:Appendix_8}
\end{equation}

\noindent which is identical with Eqn. \ref{Eqn:SF_derivation}.

The physical coordinates of the Gauss points can be expressed in terms of the nodal physical coordinates with the aid of the shape functions $N_{e}(\xi, \eta)$ as follows:

\begin{equation}
    x = \sum_{e = 1}^{8} N_{e}(\xi, \eta) x_{e} 
    \; \; \; \; \; 
    ; 
    \; \; \; \; \; 
    y = \sum_{e = 1}^{8} N_{e}(\xi\eta) y_{e} 
\end{equation}

\noindent where the subscript $e$ denotes nodal values and the shape functions are provided in Table \ref{Table:Shape_functions_quadelems}. We can now define all the terms in Eqn. \ref{Eqn:Appendix_8}:

\begin{itemize}

    \item First-order derivatives of physical w.r.t. natural coordinates (components of the element Jacobian matrix):
    
        \begin{equation}
        \dfrac{\partial x}{\partial \xi} = \sum_{e = 1}^{8} \dfrac{\partial N_{e}(\xi,\eta)}{\partial \xi} x_{e} \; \; , \; \; \mathrm{similar \ for \ other \ terms}
        \end{equation}
    
    \item First-order derivatives of non-local strain w.r.t. physical coordinates:

        \begin{equation}
        \dfrac{\partial \bar{\varepsilon}_{eq}}{\partial x} = \dfrac{\partial \bar{\varepsilon}_{eq}}{\partial \xi} \dfrac{\partial \xi}{\partial x} + \dfrac{\partial \bar{\varepsilon}_{eq}}{\partial \eta} \dfrac{\partial \eta}{\partial x} = \left(\dfrac{\partial N}{\partial \xi} \bar{\varepsilon}_{eq,e}\right) \dfrac{\partial \xi}{\partial x} + \left(\dfrac{\partial N}{\partial \eta} \bar{\varepsilon}_{eq,e}\right) \dfrac{\partial \eta}{\partial x} 
        \; \; , \; \; \mathrm{similar \ for \ other \ terms}
        \end{equation}
    
    \item Second-order derivatives of physical w.r.t. natural coordinates:
    
        \begin{equation}
        \dfrac{\partial^{2} x}{\partial {\xi}^{2}} = \sum_{e = 1}^{8} \dfrac{\partial^{2} N_{e}(\xi,\eta)}{\partial {\xi}^{2}} x_{e} \; \; , \; \; \mathrm{similar \ for \ other \ terms}
        \end{equation} 

    \item Second-order derivatives of non-local strain w.r.t. natural coordinates:
    
        \begin{equation}
        \dfrac{\partial^{2} \bar{\varepsilon}_{eq}}{\partial {\xi}^{2}} =
        \dfrac{\partial^{2} N_{e}(\xi,\eta)}{\partial {\xi}^{2}} \bar{\varepsilon}_{eq,e} \; \; , \; \; \mathrm{similar \ for \ other \ terms}
        \end{equation}
    
\end{itemize}

The second order derivatives of the shape functions with respect to the element natural coordinates are also provided in Table \ref{Table:Shape_functions_quadelems}, and the terms $\dfrac{\partial \xi}{\partial x}$, $\dfrac{\partial \eta}{\partial x}$, $\dfrac{\partial \xi}{\partial y}$ and $\dfrac{\partial \eta}{\partial y}$ are the components of the inverse element Jacobian matrix.

\begin{table}[H]
\caption{Shape functions and second-order partial derivatives for quadratic elements}
\label{Table:Shape_functions_quadelems} \centering
\begin{tabular}{ c c c c c}
    \hline
    \bf Node (e) & \bf Shape function ($N_{e}(\xi,\eta)$) & $(\partial^{2} N_{e}(\xi,\eta)) / (\partial {\xi}^{2})$ & $(\partial^{2} N_{e}(\xi,\eta)) / (\partial {\eta}^{2})$ & $(\partial^{2} N_{e}(\xi,\eta)) / (\partial \xi \partial \eta)$  \\
    \hline
    1 & $- (1 - \xi)(1 - \eta)(1 + \xi + \eta) / 4$ & $(1 - \eta) / 2$ & $(1 - \xi) / 2$ & $(1 - 2\xi - 2\eta) / 4$ \\
    2 & $- (1 + \xi)(1 - \eta)(1 - \xi + \eta) / 4$ & $(1 - \eta) / 2$ & $(1 + \xi) / 2$ & $(2\eta - 2\xi - 1) / 4$ \\
    3 & $- (1 + \xi)(1 + \eta)(1 - \xi - \eta) / 4$ & $(1 + \eta) / 2$ & $(1 + \xi) / 2$ & $(1 + 2\xi + 2\eta) / 4$ \\
    4 & $- (1 - \xi)(1 + \eta)(1 + \xi - \eta) / 4$ & $(1 + \eta) / 2$ & $(1 - \xi) / 2$ & $(2\xi - 2\eta - 1) / 4$ \\
    5 & $(1 - {\xi}^{2})(1 - \eta) / 2$ & $\eta - 1$ & $0$ & $\xi$ \\
    6 & $(1 + \xi)(1 - {\eta}^{2}) / 2$ & $0$ & $- \xi - 1$ & $- \eta$ \\
    7 & $(1 - {\xi}^{2})(1 + \eta) / 2$ & $- 1 - 2\eta $ & $0$ & $- \xi$ \\
    8 & $(1 - \xi)(1 - {\eta}^{2}) / 2$ & $0$ & $\xi - 1$ & $\eta$ \\ 
    \hline
\end{tabular}
\end{table}

\newpage
\section{I-FENN Residuals}
\label{Appendix:IFENN_Residuals}

Here we present the residuals of the models analyzed in the parametric study of Section \ref{Section:SNT_IFENN_NR}. We observe a healthy reduction in both the displacement-based (Eqn. \ref{Eqn_convergence_criterion_IFENN}) and the stress-based (Eqn. \ref{Eqn_convergence_criterion_IFENN_secondary}) residuals, across all cases (different data vs. physics combinations as well as different normalization approaches).

\begin{figure}[H]
    \centering
    \includegraphics[width=0.9\linewidth]{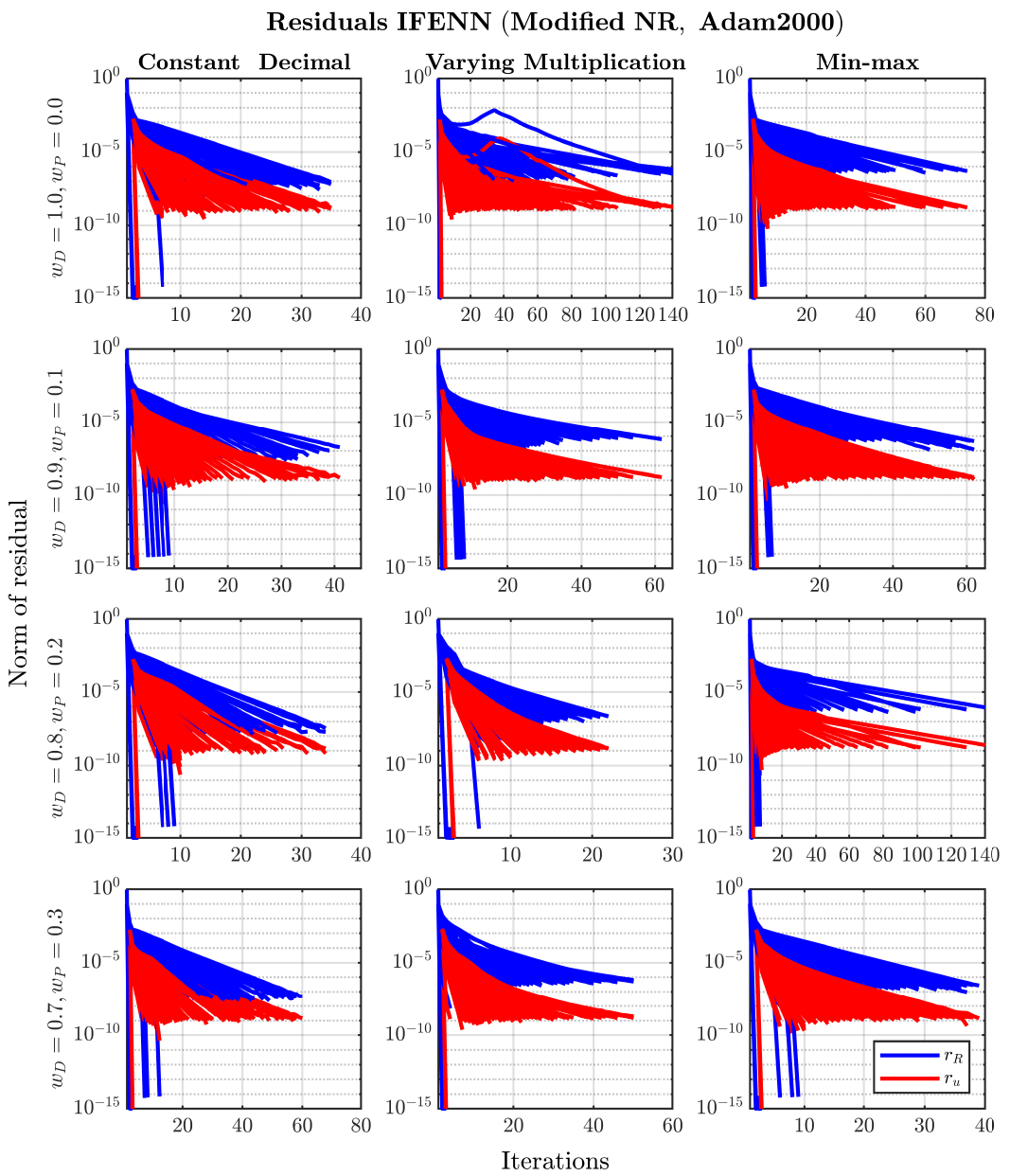}
    \caption{Minimization of the I-FENN residuals.}
    \label{Fig:Figure_Residuals_SingleNotch_IFENN_MNR_Ad2000}
\end{figure}

\end{document}